**Chemical Perspectives on Cuprate Superconductivity: Hole-Protected Spin Dimers as *d*-Wave Cooper Pairs**

Myung-Hwan Whangbo[1,*] and Reinhard K. Kremer[2]

[1] Department of Chemistry, North Carolina State University, Raleigh, NC 27695-8204, USA

[2] Max Planck Institute for Solid State Research, Heisenbergstrasse 1, D-70569 Stuttgart, Germany

M.-H. Whangbo: mike_whangbo@ncsu.edu

**Abstract**

The observation that the coherence lengths in high-$T_c$ cuprate superconductors are extremely small motivated us to search for its real-space structure by examining the consequence of hole-doping in $CuO_4$ units in high-$T_c$ cuprates. We analyzed aggregates of hole-doped $CuO_4$ units that phase-separate from hole-doped $CuO_2$ layers in terms of their segregation-pressure indices defined in this work. Our search suggests that each *d*-wave Cooper pair is an antiferromagnetically coupled spin dimer protected in an antiferromagnetic hole dimer (AHD), and that the antiferromagnetic chains of AHDs provide channels through which such spin dimers can move collectively without perturbation. Based on these suggestions and their implications, we examined the causes for several seemingly puzzling observations on high-$T_c$ cuprate superconductors to find that they are naturally explained by the suggestions. These observations include the origin of the commensurate and incommensurate charge and spin density waves, the nature of the superconducting gap $\Delta_{sc}$, the cause for the pseudogap $\Delta_{ps}$ and the linear decrease of the pseudogap transition temperature $T^*$ on hole density $p$, the reason for the near-IR absorption gap $\Delta_{nir}$ associated with the midpoint of the nearly flat near-IR absorption, the reason for the strange metallic state above $T_c$, and the limitation of superconductivity to a rather narrow doping concentration $p$. The energy gaps derived in our study are consistent with the characteristic energies observed for high-$T_c$ cuprates. Our findings provide real-space support for Anderson's conjecture that spin exchange is the pairing mechanism for the *d*-wave Cooper pairs of high-$T_c$ cuprates.

## 1. Introduction

Four decades ago, Bednorz and Müller discovered the first high-$T_c$ superconductor in the Ba-La-Cu-O perovskite.[1] Their landmark discovery launched an intense and unprecedented search for new superconducting materials with increasingly higher critical temperatures. Since then, several new families of superconductors have been identified. To date, however, high-$T_c$ cuprates with perovskite-derived crystal structures remain the only superconductors with critical temperatures above the boiling point of liquid nitrogen under ambient conditions, making them promising for commercial applications. The discovery of high-$T_c$ superconductors not only inspired extensive experimental work, e.g., thin film technologies of oxides, but also profoundly influenced theories of strongly correlated systems with unconventional ground states. Despite extensive experimental and theoretical work, however, many aspects of the mechanism underlying superconductivity remain unresolved. In particular, the driving force responsible for electron pairing and lossless charge transport below $T_c$ is still not fully understood.

Over the past four decades, several characteristic features of high-$T_c$ cuprate superconductors have been identified:

(1) All high-$T_c$ cuprates contain $CuO_2$ square-planar layers, which may exhibit buckling and lowered symmetry rather than remaining perfectly flat. Such quasi-two-dimensional crystal structures are a defining feature of nearly all superconductors discovered to date.

(2) In most high-$T_c$ superconductors, superconductivity arises from hole doping in the $CuO_2$ layers; electron-doped systems are rather rare.

(3) The parent compounds of high-$T_c$ cuprates are antiferromagnetic insulators with spin exchange (“superexchange”) energies of about 100 meV, equivalent to roughly 1200 K. Doping via interlayer charge reservoirs rapidly suppresses antiferromagnetic order.

(4) High-$T_c$ superconductivity is confined to a narrow doping level, typically $0.2 < p < 0.3$ holes per Cu ion.

(5) The $T_c$ phase boundary forms an "inverted parabolic dome," occasionally with boundary indentations,

(6) The superconducting gap exhibits *d*-wave symmetry.

(7) Upon hole doping, a precursor "pseudogap" phase emerges and persists at relatively high temperatures, comparable to or higher than the Néel temperature, with its phase boundary extending into or touching the superconducting dome. The pseudogap phase is often attributed to the formation of local hole pairs, namely, short-range electron-hole bound states believed to emerge in the under-doped regime at high temperatures above $T_c$.

(8) In the overdoped regime ($p > p_{max}$), high-$T_c$ cuprates enter a Fermi-liquid phase, although spin fluctuations and strong electron correlations appear to remain.

(9) Commensurate and incommensurate charge- and spin-ordering phenomena ("stripes") emerge as precursors to or competing with superconductivity. Increasing evidence also points to microscopic inhomogeneities, including mesoscopic phase segregation.

(10) Further doping above the superconducting dome drives the pseudogap phase into a "strange metal" phase, marked by anomalous electronic transport such as the nearly linear temperature dependence of the in-plane resistivity over a broad temperature range. This phase approaches, and may extend into, the superconducting region by lowering the temperature.

In summary, at high-temperatures the high-$T_c$ superconductors are dominated by an antiferromagnetic, spin-ordered Néel phase that is rapidly suppressed by doping. With increasing doping, the pseudogap phase and subsequently the strange-metal phase emerge, while superconductivity develops at lower temperatures. Additional charge- and spin-density phases appear before superconductivity occurs within a relatively narrow doping range. *T-p* phase diagrams for high-$T_c$ superconductors summarize the key features outlined above and show how the material phases vary with temperature *T* and hole concentration *p*.[2,3]

In this work, we take a chemical perspective to identify the likely real-space structure of *d*-wave Cooper pairs in cuprate superconductors. This approach was inspired by Watson and Crick's path to the DNA double helix.[4] Starting from the four nitrogenous bases—adenine (A), thymine (T), guanine (G), and cytosine (C)—they looked for bound base pairs of similar shape that could form rungs between two spiraling sugar-phosphate chains. The chemical mechanism linking the bases was initially unknown, but the recognition of hydrogen bonding as the mechanism quickly established G-C and T-A as the only pairs suitable for building the complementary double helix. By analogy, our problem has three parts: identifying the real-space structural units that accommodate Cooper pairs, determining the mechanism that binds two electrons into such pairs, and establishing how these units must be arranged to produce superconductivity.

Our work was motivated by the short coherence length ξ (i.e., the very small size) of the *d*-wave Cooper pair, which is approximately 1.5–3.8 nm,[5-8] although larger ξ values of ~7 and ~11 nm were also reported.[7] The shorter range corresponds roughly to 4*a*–9*a*, where *a* is the repeat distance, or the linear Cu–O–Cu bond length, of the ideal hole-free $CuO_2$ layer. This short length suggests that the structure and shape of a *d*-wave Cooper pair can be identifiable by analyzing the structure of a hole-doped $CuO_2$ layer in terms of hole-doped and hole-free $CuO_4$ units. Furthermore, according to the Heisenberg uncertainty principle, greater precision in real space entails lower precision in reciprocal space. Thus, once the probable real-space size and shape of the *d*-wave Cooper pair are identified, we thought it more fruitful to address various puzzling observations in cuprate superconductors from a real-space perspective rather than from a reciprocal-space perspective.

Our work was guided by the implications of three seminal observations: (1) Anderson's conjecture that spin-exchange-coupled electrons must be accepted as the Cooper pairs in cuprate superconductors,[9] (2) the realization[10-29] that electronic phase separation occurs in doped $CuO_2$ layers, and (3) Wen et al.'s finding from their high-field X-ray scattering study[30] of $La_{1.885}Sr_{0.115}CuO_4$ (LSCO) that charge density wave (CDW), spin density wave (SDW), and superconductivity coexist. Since the implications of these observations are crucial in the development of our work, we summarize them below:

(1) In his 2016 "Last Words on the Cuprates," Anderson pointed out that the dominant interaction coupling superconducting electron pairs is the same spin-exchange interaction that drives undoped cuprates into an antiferromagnetic (AFM) Mott-insulating state.[9] This conjecture has two immediate implications. (a1) AFM spin dimers of $S = 1/2$ magnetic ions, in which two electrons are bound by spin exchange, are natural candidates for Cooper pairs. This reasoning further implies that (a1-1) the antiferromagnetically-coupled state in these spin dimers must be protected from surrounding spins, suggesting that they are "protected" AFM spin dimers (protected spin dimers, for short), and (a1-2) these protected spin dimers must form protected channels through which they can move to generate superconducting current. (a2) The three characteristic energy gaps of cuprate superconductors—namely, the $\Delta_{pg}$[31-46] of the pseudogap state, the $\Delta_{sc}$[45] of the superconducting state, and the $\Delta_{nir}$[48-51] associated with the center of the nearly flat near-IR absorption peak—must therefore be related to the nearest-neighbor spin exchange $J$, with relative magnitudes decreasing in the order $\Delta_{nir} >> \Delta_{pg} > \Delta_{sc}$.

(2) Already in their 1994 review,[16] Kivelson and Emery emphasized the importance of electronic phase separation, in which holes segregate into hole-rich and hole-poor regions, thereby affecting superconductivity and charge ordering. Electronic phase separation has several implications. (b1) The building blocks of a doped $CuO_2$ layer are $CuO_4$ units. Chemically, the hole-rich regions contain more hole-doped than hole-free $CuO_4$ units, whereas the opposite is true in the hole-poor regions. (b2) Because the doping concentration $p$ in high-$T_c$ cuprates is typically below 0.3,[2,3] hole-free $CuO_4$ units constitute the majority phase and hole-doped $CuO_4$ units the minority phase in a hole-doped $CuO_2$ layer. In heterogeneous two-component systems, phase segregation generally takes place to reduce contact between the two components by forming larger aggregates of the minority component. Thus, (b2-1) aggregates of hole-doped $CuO_4$ units with different sizes and shapes are expected to occur in a hole-doped $CuO_2$ layer, even if they are too small to be detected by diffraction and scattering methods, and (b2-2) the charge and spin ordering observed in the stripe phase[52-60] is most likely associated with such aggregates.

(3) In their 2023 high-field X-ray scattering study of $La_{1.885}Sr_{0.115}CuO_4$ (LSCO), Wen et al.[30] showed that (c1) at low temperatures, LSCO contains two distinct CDW phases: a majority short-range CDW phase that coexists with superconductivity and a minority longer-range CDW phase that coexists with static SDW; (c2) in the longer-range CDW phase, where the CDW is commensurate with the SDW, doped $CuO_2$ layers contain two types of stripe regions, one with long charge-ordered stripes aligned along the *a*-direction ($||a$) and the other with stripes aligned along the *b*-direction ($||b$); and (c3) in the short-range CDW phase, "square" segments of short $||a$-stripes and those of $||b$-stripes pack together orthogonally to cover the superconducting region, such that each $||a$-segment is surrounded by four ||*b*-segments to form nodes, and vice versa, at the four corners of each segment. These findings led us to expect that both the *d*-wave Cooper pairs and the superconducting mechanism of cuprate superconductors can be understood once the real-space structures responsible for the CDW and SDW features of cuprate superconductors are identified.

In what follows, our work is organized as follows: Section 2 provides a brief description of how we analyzed the phase-segregation of hole-doped $CuO_2$ layers. In Section 3, we define "protected" hole-doped $CuO_4$ units and their aggregates, introduce their segregation-pressure indices, and examine their phase segregation from the region of hole-free $CuO_4$ units quantitatively based on those indices. In Section 4, we show that the CDW and SDW features observed for the stripes, both commensurate and incommensurate, are naturally explained by AFM chains made up of protected hole-doped $CuO_4$ units, which are referred to there as antiferromagnetic hole-dimer (AHD) chains; this leads us to conclude that the stripes are composed of AHD chains. Wen et al.'s observation (c3) enabled us to conclude that short AHD chains are responsible for the superconductivity of cuprate superconductors. From this conclusion together with our finding that the AHD allow superconducting current flow only along the chain direction compelled us to conclude that the nodes between ||*a*- and ||*b*-segments of short AHD chains must act as Josephson junctions,[61] allowing superconducting current to flow in all directions in the hole-doped $CuO_2$ layers. In Section 5, we examine the origin of the three characteristic energy gaps observed in cuprate superconductors solely based on spin-exchange interactions. In Section 6, we summarize our principal findings and address three important issues: why singlet-state spin dimers behave as bosonic entities that can move freely through the protected channels of AHD chains; the possible origins of charge- and spin-order periods other than 3*a* and 6*a*, respectively; and the probability that AHD-based hole aggregates can are the antiferromagnetic puddles needed to account for the linear resistivity $\rho \propto T$ in the strange-metal state. Finally, we present our conclusions in Section 7.

## 2. Experimental section

Our study analyzes experimental results reported in the open literature. In searching for the relevant information and the associated references from the literature accumulated over the past four decades of intense research efforts, the AI, Wikipedia, and WWW searches have been indispensable for our study. The hole-doped $CuO_2$ layers of cuprate superconductors are heterogeneous, comprising both hole-free and hole-doped $CuO_4$ units. In these layers, the hole-doped units cluster into aggregates that segregate from hole-free regions. To quantify the aggregate growths, we define "protected holes" as hole-doped $CuO_4$ units encircled by eight hole-free $CuO_4$ units and use them as the building blocks of aggregates and defined their segregation-pressure indices to characterize their driving force to grow larger. Further details are given in the next section.

## 3. Heterogeneity of hole-doped $CuO_2$ layers and their phase-segregation

### 3.1. Hole-doped and hole-free $CuO_4$ units

Cuprates contain $CuO_2$ layers (**Fig. 1a**) built from corner-sharing square-planar $CuO_4$ units (**Fig. 1b**). Charge neutrality is generally maintained by cations lying in between these layers. When charge balance places all Cu atoms in the +2-oxidation state, the $Cu^{2+}$ ions ($d^9$, S = 1/2) in each $CuO_2$ layer couple antiferromagnetically, producing a magnetic insulating state. If charge balance instead gives Cu an average oxidation state of $+(2 + p)$, the $CuO_2$ layer is hole-doped with hole concentration $p$. Hole doping removes electrons from the highest occupied d-state region of the hole-free $CuO_2$ layer, located near the wave-vector point M = ($a^*/2$, $b^*/2$) in the electronic band structure.[62] This interpretation is consistent with the observation of hole pockets centered at M in underdoped $YBa_2Cu_3O_{6+x}$.[63,64] At M, each $CuO_4$ unit is described by its magnetic orbital: the singly occupied state formed by σ-antibonding interaction between the Cu $x^2{-}y^2$ orbital and the 2p orbitals of the four surrounding O atoms (**Fig. 1c**),[65] which gives rise to the magnetic properties of $Cu^{2+}$ ions. Doping a hole into a $CuO_4$ unit is therefore equivalent to removing the electron from this magnetic orbital. This removal eliminates σ-antibonding in the Cu–O bonds associated with occupation of the magnetic orbital, thereby shortening the Cu–O bonds.[66,67] In terms of orbital interactions,[68] the loss of σ-antibonding can be viewed as arising from the formation of σ-bonding interactions between the Cu $x^2{-}y^2$ orbital of a hole-doped $CuO_4$ unit and the 2p orbitals of the four O atoms, as shown in **Fig. 1d**.

In short, hole-doped $CuO_4$ units differ from hole-free units in two keyways: they have shorter Cu–O bonds and are more electron deficient. As noted above in point (b2-1), hole-doped $CuO_4$ units are therefore expected to form aggregates of various sizes and shapes in a hole-doped $CuO_2$ layer. These aggregates are referred to hereafter as hole aggregates. In the next section, we describe this phase segregation analytically by defining the segregation-pressure index, $I_{sp}$.

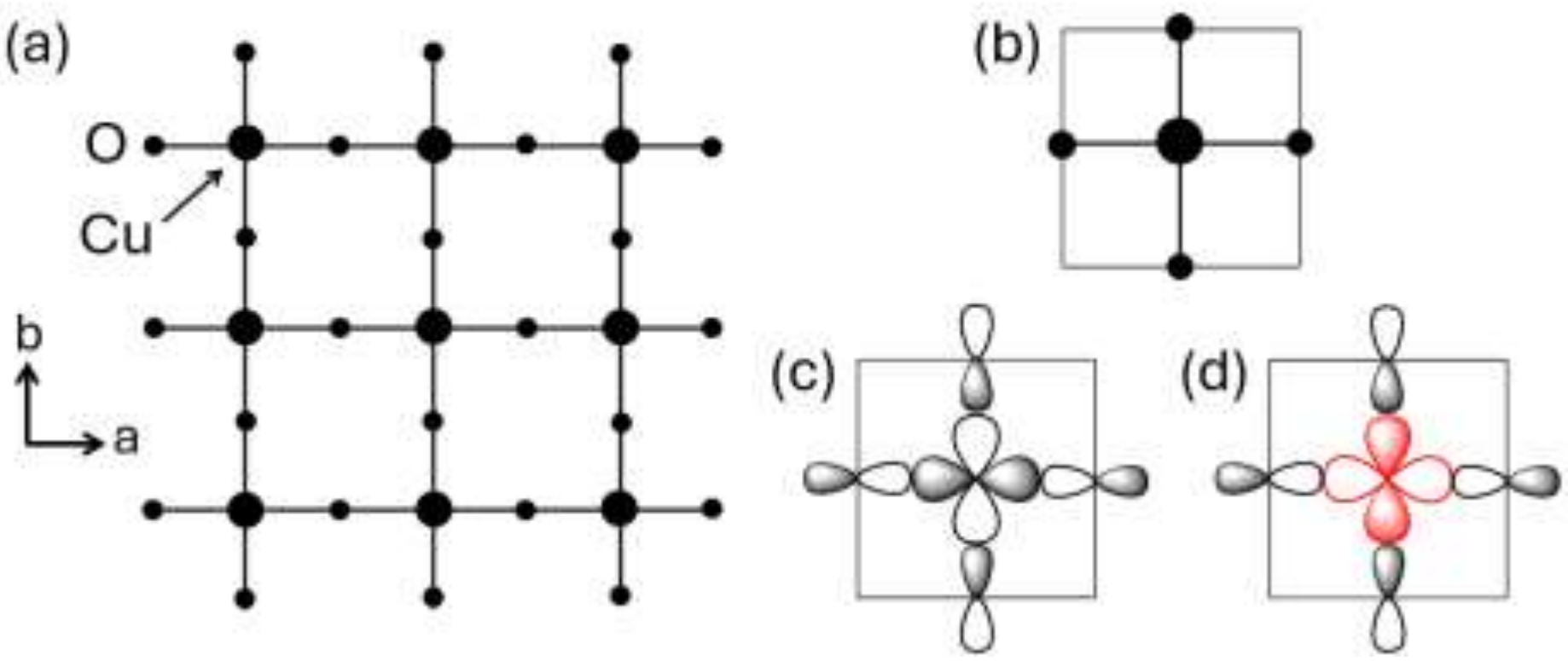

Fig. 1. (a) A schematic view of a hole-free $CuO_2$ layer. (b) Hole-free $CuO_4$ unit. (c) σ-antibonding interactions of the Cu $x^2$-$y^2$ orbital with O 2p orbitals in a hole-free $CuO_4$ unit. (d) σ-bonding interactions of the Cu $x^2$-$y^2$ orbital with O 2p orbitals in a hole-doped $CuO_4$ unit.

### 3.2. Segregation-pressure index and hole aggregates

#### 3.2.1. Simplified representations of hole-doped and hole-free $CuO_4$ units

To facilitate the discussion below, we introduce simplified representations of hole-free and hole-doped $CuO_4$ units, shown in **Figs. 2a** and **2b**, respectively. These representations provide a compact way to describe extended arrangements. For example, a linear trimer of three "edge-sharing" hole-free $CuO_4$ units (**Fig. 2c**, top) is represented schematically in the lower part of **Fig. 2c**. (Here the term "edge-sharing" was used to emphasize the square box containing a $CuO_4$ unit and its Cu $x^2$-$y^2$ orbital. From the viewpoint of chemical bond formation between adjacent $CuO_4$ units, it is conventional to use "corner-sharing".) Similarly, a linear trimer in which one hole-doped $CuO_4$ unit shares edges with two hole-free $CuO_4$ units (**Fig. 2d**, top) is represented schematically in the lower part of **Fig. 2d**.

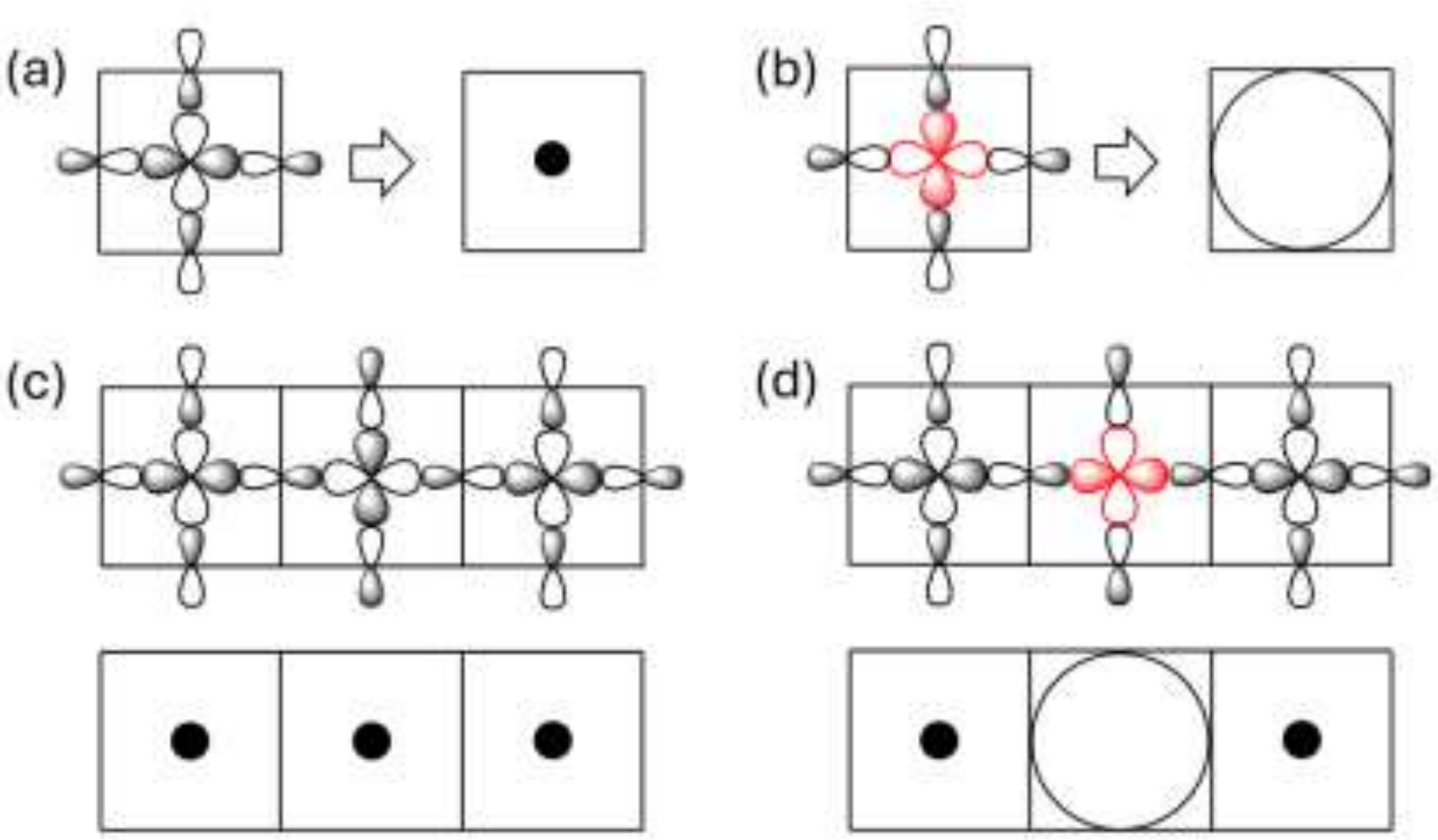

Fig. 2. (a) Simplified notation (right) for hole-free $CuO_4$ unit (left). (b) Simplified notation (right) for hole-free $CuO_4$ unit (left). (c) Simplified notation (bottom) for three hole-free $CuO_4$ units (top). (d) Simplified notation (bottom) for hole-doped $CuO_4$ unit sandwiched between two hole-free $CuO_4$ units (top).

### 3.2.2. Protected holes

Using this notation, **Fig. 3a** shows a 3 × 3 array of nine edge-sharing hole-free $CuO_4$ units, whereas **Fig. 3b** shows the corresponding array with the central $CuO_4$ unit hole-doped. Because a hole-free $CuO_2$ layer has an AFM spin arrangement, two hole-doped structures must be distinguished. Removing a down-spin electron produces an "up-spin hole monomer" (↑-HM; **Fig. 3c**, left), whereas removing an up-spin electron produces a "down-spin hole monomer" (↓-HM; **Fig. 3d**, left). In both cases, the central hole-doped $CuO_4$ unit is surrounded by a ring of eight antiferromagnetically coupled hole-free $CuO_4$ units, so that these units carry no net spin moment. We use the terms "↑-HM" and "↓-HM" to emphasize the point that the four nearest-neighbor $CuO_4$ units of the hole doped $CuO_4$ carry up-spins in the ↑-HM but down spins in the ↓-HM, thereby allowing them to couple antiferromagnetically and form hole aggregates of various sizes.

These HMs can be further simplified by omitting the small square boxes representing individual $CuO_4$ units, as shown in the right panels of **Figs. 3c** and **3d**. We use the ↑-HM and ↓-HM as building blocks for the hole aggregates discussed below. Note that each HM represents a "protected hole-doped $CuO_4$ unit" in that a hole-doped $CuO_4$ unit is surrounded by a shell of eight antiferromagnetically coupled hole-free $CuO_4$ units.

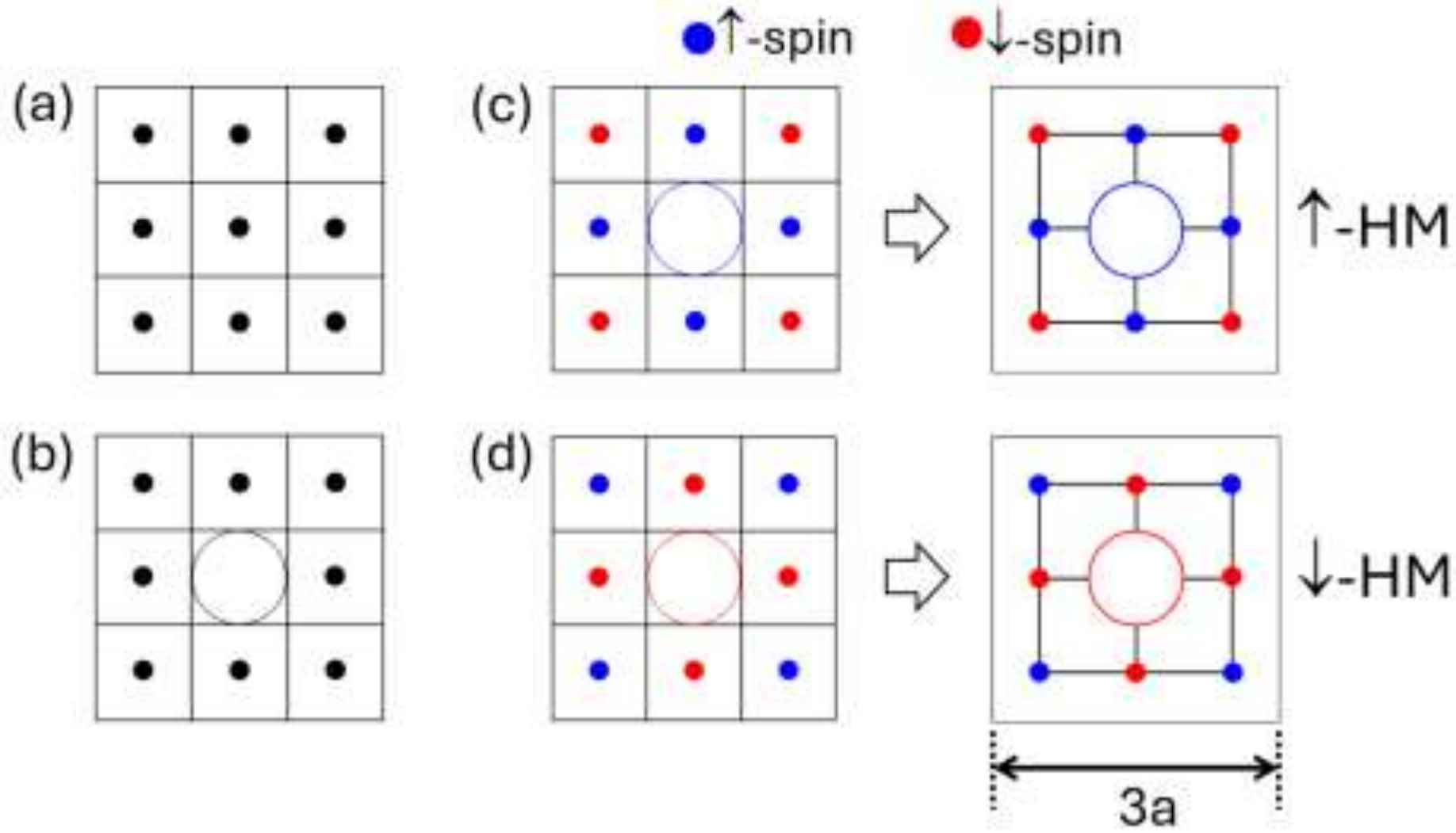

Fig. 3. (a) 3×3 array of hole-free $CuO_4$ units. (b) Hole-doped $CuO_4$ unit surrounded by eight hole-free $CuO_4$ units. (c) Simplified representation (right) of "up-spin" hole monomer (right) in which a hole-doped $CuO_4$ unit is generated by removing a down-spin (left). (d) Simplified representation (right) of "down-spin" hole monomer (right) in which a hole-doped $CuO_4$ unit is generated by removing an up-spin (left).

**3.2.3. Segregation-pressure index and phase segregation**

In this section, we examine implications (b1) and (b2), together with the related points (b2-1) and (b2-2), which follow from the observation of electronic phase separation.[10-29] To explore these implications, we must describe how the sizes and shapes of hole aggregates depend on the doping concentration $p$ and temperature $T$. This requires a quantitative measure for the extent of contact between a hole aggregate and the surrounding hole-free phase. For this purpose, we define the segregation-pressure index, $I_{sp}$, for a given hole aggregate.

Consider a hole aggregate containing $n_h$ holes and having a peripheral edge of length $L_{pe}$. When this aggregate is embedded in a region of hole-free $CuO_4$ units, the extent of its contact with the hole-free phase is measured by $L_{pe}$. To express the segregation pressure per hole, we define $I_{sp}$ as

$$I_{sp} = \frac{L_{pe}}{n_h} \tag{1}$$

Thus, a hole aggregate with a large $I_{sp}$ is under stronger pressure to transform into an aggregate with a smaller $I_{sp}$. To illustrate this point, we construct several hole aggregates by antiferromagnetically coupling up-spin and down-spin HMs (**Figs. 3c** and **3d**). Starting from the HM, structure (1), we consider the larger aggregates (2)–(5) described below:

(2) AFM hole dimer (AHD; **Fig. 4a**)

(3) AHD chain of *m* AHDs (**Fig. 4b**)

(4) Stripe consisting of *k* AHD chains, each containing *m* AHDs (**Fig. 4c**), hereafter denoted as *k*-stripe. A 1-stripe is equivalent to an AHD chain.

(5) A large square (LS) aggregate obtained from the *k*-stripe as both *k* and *m* become large.

The $I_{sp}$ values for structures (1)–(5) are summarized in **Table 1** and decrease in the order (1) > (2) > (3) > (4) > (5).

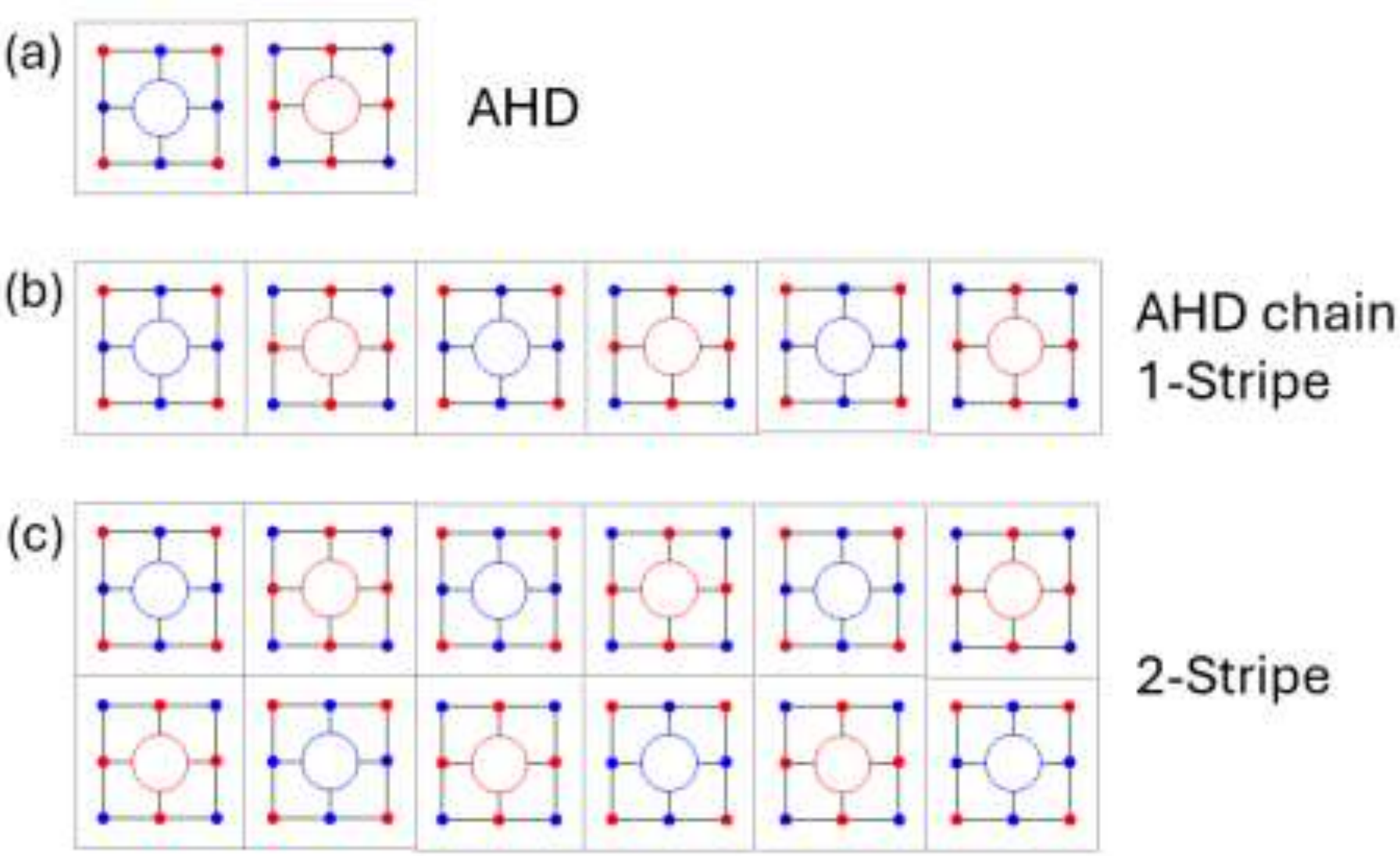

Fig. 4. Hole-aggregates obtained by antiferromagnetically coupling ↑-HM and ↓-HM units: (a) Antiferromagnetic hole dimer (AHD). (b) Antiferromagnetic chain of AHD units ($m$ = 3), which is equivalent to 1-stripe. (c) 2-stripe ($m$ = 3).

Table 1. Segregation-pressure indices of some hole-aggregates, where $m$ is the number of AHDs.

| Hole-aggregate | $I_{sp}$ (in $a/h$) |
|---|---|
| HM | 12 |
| AHD | 9 |
| AHD chain | $6 + 3/m$ |
| $k$-stripe | $6/k + 3/m$ |
| Large square | $6/m$ |

From **Table 1**, the decrease in the segregation pressure $\Delta I_{sp}$ (in units of $a/h$) associated with a structural change from a small to a large hole aggregate can be summarized as:

$\Delta I_{sp}$ = 3 from HM to AHD

$\Delta I_{sp}$ = 3 from AHD to 1-stripe ($m \rightarrow \infty$)

$\Delta I_{sp}$ = 3 from 1-stripe ($m \rightarrow \infty$) to LS

The $\Delta I_{sp}(k)$ associated with a structural change from $k$-stripe to ($k$+1)-stripe is written as

$$\Delta I_{sp}(k) = \frac{6}{k(k+1)}. \quad (2)$$

This quantity equals 3 at $k = 1$ and 1 at $k = 2$, then decreases monotonically with increasing $k$. Thus, the driving force for further growth weakens as a $k$-stripe widens, making the formation of very wide $k$-stripes—and hence large square hole aggregates (**Fig. 5a**)—unlikely through stripe growth.

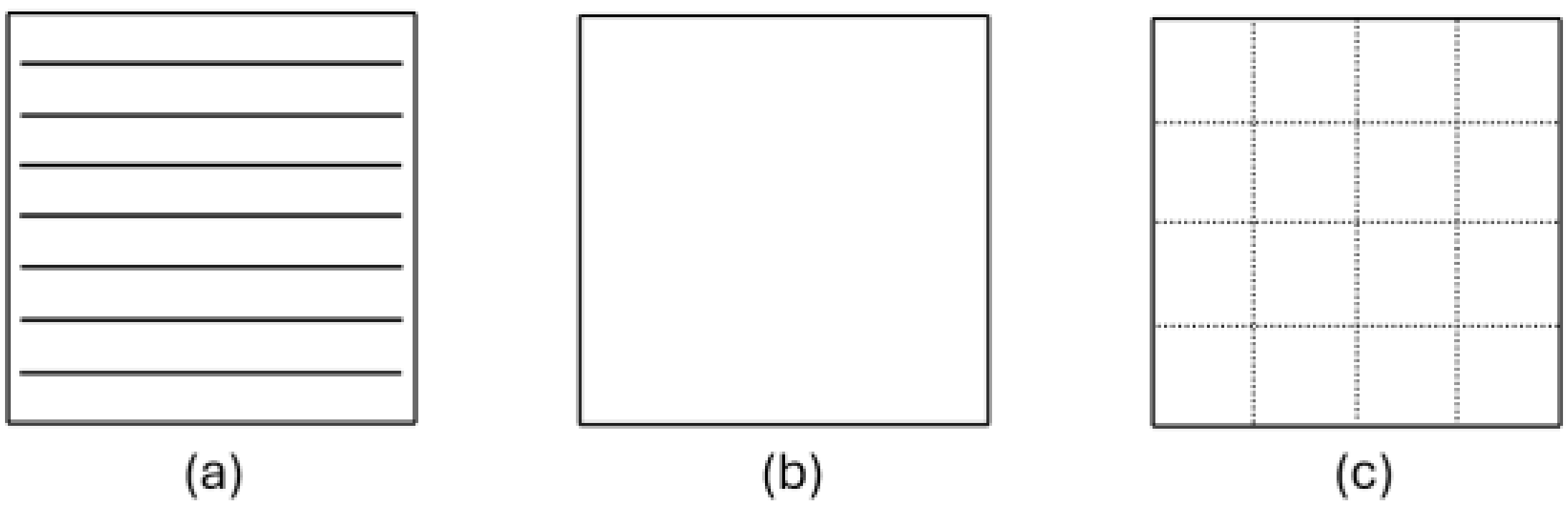

Fig. 5. (a) Large square aggregate formed from a very wide $k$-stripe. (b) Large square aggregate formed from a small square aggregate. (c) Assembly of small square-shaped aggregates.

An efficient way to reduce the segregation pressure of hole aggregates is to form square aggregates from ↑-HM and ↓-HM units, as illustrated in **Fig. 6**. For a square aggregate with side length $3qa$, where $q$ is the number of HMs along each side, the peripheral edge length is $L_{pe} = 12qa$, and the number of holes is $q^2$. In units of $a/h$, the segregation-pressure index $I_{sp}(q)$ is therefore

$$I_{sp}(q) = \frac{12}{q} \qquad (3)$$

This index decreases rapidly with $q$: $I_{sp}(1) = 12$, $I_{sp}(2) = 6$, and $I_{sp}(3) = 4$. The decrease in segregation pressure, $\Delta I_{sp}(q)$, associated with growth from a $q$-square to a $(q + 1)$-square is given by

$$\Delta I_{sp}(q) = \frac{12}{q(q+1)}. \qquad (4)$$

The driving force toward the next larger aggregate is strongest at $q = 1$, where $\Delta I_{sp}(1) = 6$, but it decreases rapidly as $q$ increases; for example, $\Delta I_{sp}(10) = 0.11$. Here, $q$ refers to the number of HM units along a side, not the number of HDs. This square-growth mode is therefore unlikely to produce a uniform large square aggregate (**Fig. 5b**). Instead, many small square segments are more likely to form because they provide a larger reduction in $\Delta I_{sp}(q)$. In summary, at higher hole densities, square growth is statistically more probable than linear growth. If these small-square segments assemble into a compact structure with a short peripheral edge (**Fig. 5c**), such a fragmented square aggregate should remain entropically favorable. This point becomes important in Section 4.

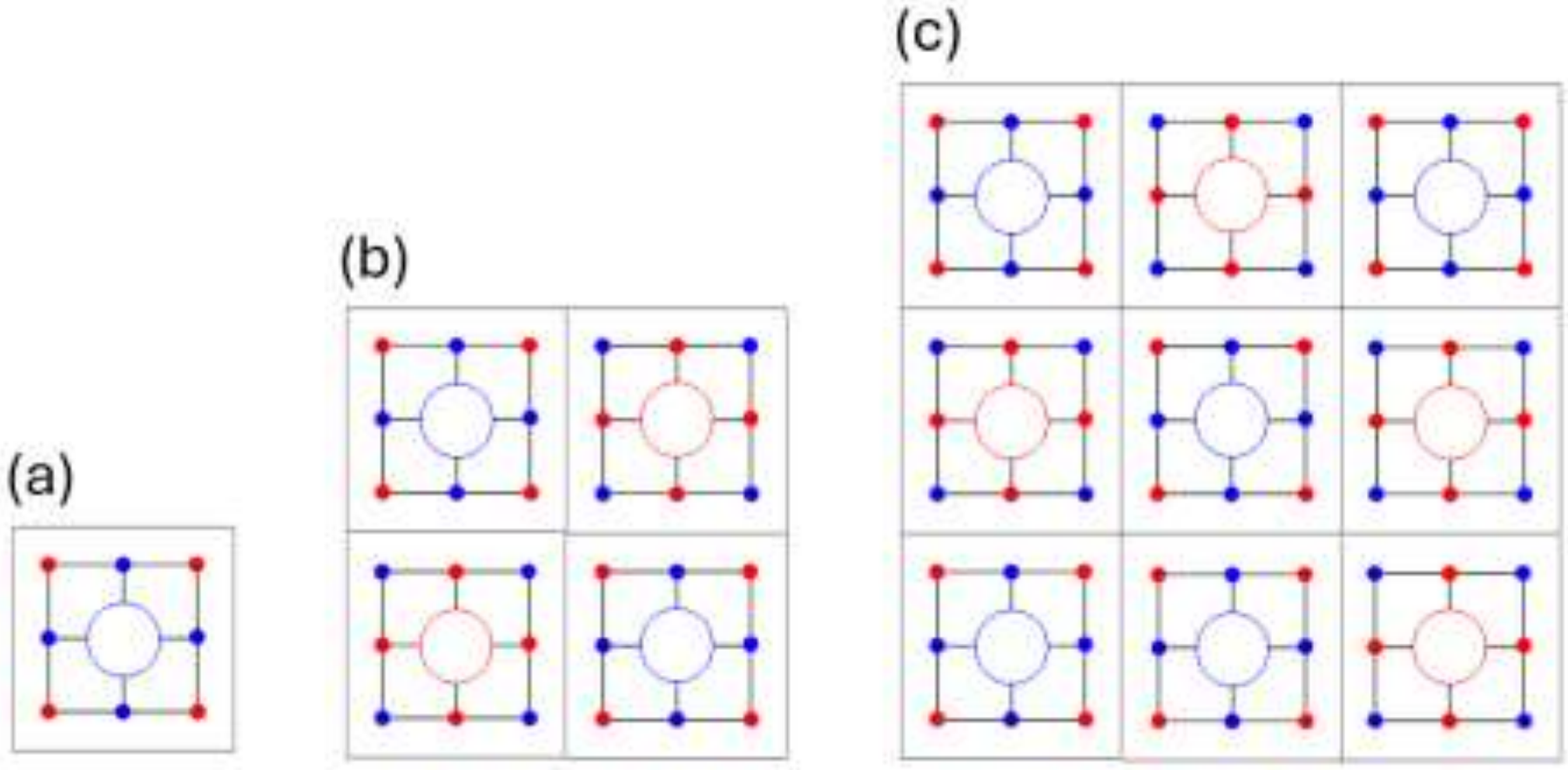

Fig. 6. Square-growth of hole-aggregates: (a) HM unit. (b) 2×2 aggregate of four HM units. (c) 3×3 aggregate of nine HM units.

### 3.3. Nature of phase segregation in homologous series of hole aggregates

We now examine the segregation of hole aggregates from hole-free $CuO_4$ regions in terms of entropy and the associated free-energy change. We begin with the segregation-pressure index $I_{sp}(m)_1$ for the 1-stripe homologous series:

$$I_{sp}(\mathrm{m})_1 = 6 + \frac{3}{m} \qquad (5)$$

For $m > 5$, the $3/m$ term is already less than 10% of the first term, 6, and it becomes negligible when $m > 10$. Now consider growth from a smaller to a larger member of the series within a region of hole-free $CuO_4$ units. Because the larger aggregate is more ordered, the smaller and larger aggregates have higher and lower entropies, $S_H$ and $S_L$, respectively. At a given temperature $T_H$, the free-energy change $\Delta G_H$ required for this reorganization is

$$\Delta G_H = -T_H \Delta S = -T_H (S_L - S_H) \qquad (6)$$

When the temperature is lowered to $T_L$, the free energy of the smaller aggregate changes to

$$\Delta G_L = -S_H \Delta T = -S_H (T_L - T_H) \qquad (7)$$

If the free-energy change $\Delta G_L$ caused by lowering the temperature becomes equal to the free energy $\Delta G_H$ required for the smaller aggregate to reorganize into the larger one, we obtain

$$\Delta T = \frac{T_H}{S_H} \Delta S \qquad (8)$$

The total entropy change $\Delta S$ associated with this reorganization should scale with the extent of contact between the aggregate and the hole-free region, which is measured by the peripheral edge length $L_{pe}$. Since $L_{pe} = n_h I_{sp}$, we have $\Delta S \propto n_h I_{sp}$. Equation (5) can therefore be rewritten as

$$\Delta T \propto \left(\frac{I_{sp}}{S_H}\right) n_h \qquad (9)$$

This relation shows that the temperature decrease required for a smaller aggregate to grow into a larger member of the homologous series is proportional to the number of holes in the smaller aggregate.

For comparison, the segregation-pressure indices $I_{sp}(m)_2$ for members of the 2-stripe series are given by

$$I_{sp}(m)_2 = 3 + \frac{3}{m} \qquad (10)$$

This value is approximately half of $I_{sp}(m)_1$. However, the entropy $S_H$ is also expected to be smaller for a 2-stripe than for a 1-stripe, so the ratio $I_{sp}/S_H$ should remain comparable for the two series. Accordingly, the members of the 2-stripe series should also follow Eq. (6). The same reasoning applies to other $k$-stripe homologous series.

The linear relationship between $\Delta T$ and $n_h$ applies to homologous structures formed by linear growth, such as *k*-stripes. In contrast, this relationship is not expected for square aggregates produced by square growth, because the segregation-pressure index of each member, $I_{sp}(q)$, continues to decrease as $q$ increases.

The linear relationship between $\Delta T$ and $n_h$ for *k*-stripes formed by linear growth explains the experimentally observed linear decrease of the pseudogap transition temperature $T^*$ with increasing hole concentration $p$ in cuprate superconductors.[2,3,34] Two important implications follow. (d1) In the pseudogap state, the hole aggregates are most likely individual members of the *k*-stripe homologous series, whose lengths are expected to increase as the temperature decreases. (d2) These aggregates arise from AFM coupling between adjacent ↑-HD and ↓-HD units. Therefore, the associated energy gap is not the superconducting gap $\Delta_{sc}$,[47] but the pseudogap $\Delta_{ps}$, which represents the energy gain from AFM coupling between ↑-HD and ↓-HD units and is therefore related to the AFM exchange constant $J$. We examine the magnitude of $\Delta_{ps}$ in Section 5.

## 4. Hole aggregates responsible for charge/spin order and superconductivity

This section identifies the hole aggregates in doped $CuO_2$ layers that are relevant to charge/spin order and *d*-wave superconductivity in cuprate superconductors and discusses their structural features.

### 4.1. Characteristic features of CDW and SDW

The AHD (**Fig. 4a**) serves as the building block of the AHD chain (**Fig. 4b**). Each AHD chain exhibits a CDW with a period of $3a$ and an SDW with a period of $6a$ (**Fig. 7a**). These periodicities match those of the commensurate CDW (CCDW) observed in the stripe phase, whose charge-order pattern is described by the reciprocal-space wave vector $q_{ccdw} \approx 0.33$.[69] This wave vector corresponds to a real-space repeat length of $3a$, as predicted for the AHD chain. This agreement strongly suggests that stripes are composed of AHD chains. If so, the incommensurate CDW (ICDW) and incommensurate SDW (ISDW) observed in the stripe phase should also be describable in terms of AHD chains. The ICDW wave vectors are typically close to 0.3, i.e., $q_{icdw} \approx 0.3$,[70] corresponding to a real-space period of ~$3.3a$ and a supercell length of ~$20a$. To identify the structural origin of this incommensurability, we consider the simplest likely defect: linear antiferromagnetically-coupled trimers inserted between two HM units. When inserted between adjacent HM units, these trimers act as structural defects. One linear trimer—or any odd number of linear trimers—can occur between two ↑-HMs (**Fig. 7b**) or between two ↓-HMs, because the spin arrangement of the resulting hole-doped aggregate should remain compatible with the AFM background of the hole-free majority phase. This type of *defect* forces each ↑-HM to couple only with another ↑-HM, and each ↓-HM only with another ↓-HM. By contrast, defects containing an even number of linear trimers enable AFM coupling between a ↑-HM and a ↓-HM. The smallest such defect contains two linear trimers (**Fig. 7c**). If one in every six AHDs contains this defect, the real-space supercell length becomes $20a$, or $3.33a$ relative to the defect-free unit cell. In reciprocal-space, this gives $q_{icdw} \approx 0.3$ for the ICDW.[70] The associated ISDW has a supercell length of $40a$, giving $q_{isdw} = q_{icdw}/2$. More generally, when ICDW and ISDW coexist, $q_{icdw} \approx 2q_{isdw}$; similarly, when CDW and SDW coexist, $q_{cdw} \approx 2q_{sdw}$. This relationship has been observed experimentally.[71-77]

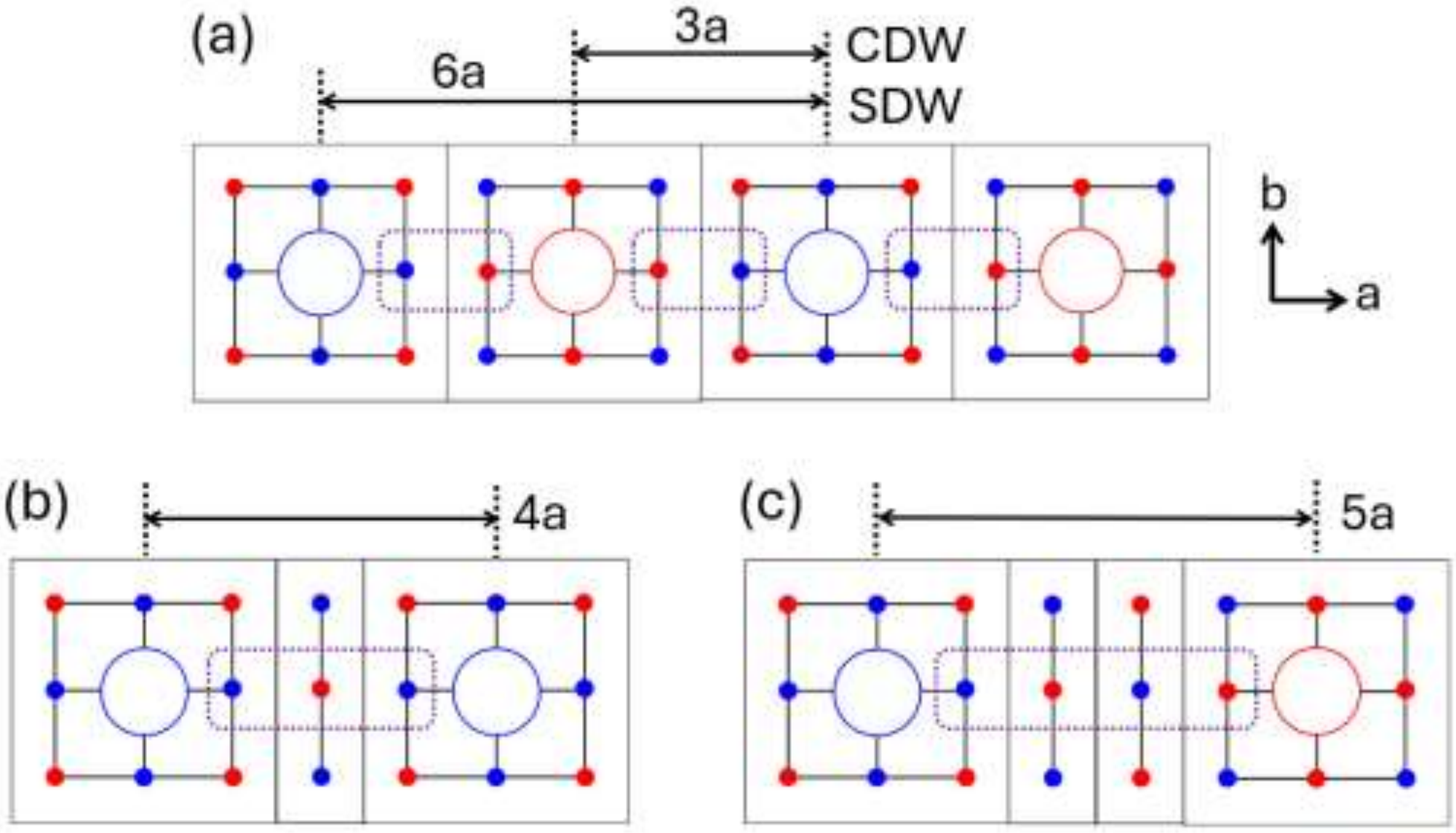

Fig. 7. (a) CDW and SDW character of an AHD chain, where the protected spin dimers are encircled by dotted lines. (b) A HD with one linear-trimer as the kink between two ↑-HMs. (c) A HD with two linear-trimers as the kink between ↑-HM and ↓-HM units.

These considerations strongly suggest that each stripe in the stripe phase is composed of AHD chains. According to the observation (c3) of Wen et al.'s study,[30] the short stripes present in the superconducting phase of LSCO are composed of short AHD chains. We therefore examine the distinctive features of AHD and AHD chains in greater detail in the following section. The possible relevance of the defect-modified AHDs such as those depicted in **Fig. 7b** and **7c** in describing the charge and/or spin stripes of periods other than 3*a* and 6*a* in several cuprates will be briefly discussed in Section 6.

### 4.2. Features of the AHD chain concerning *d*-wave superconductivity

As encircled in **Fig. 7a**, each AHD unit contains an AFM spin dimer located between two hole-sites. This spin dimer is "protected" because it cannot engage in spin-exchange interactions with other spins in the surrounding region of hole-free $CuO_4$ units. The length of the protected spin dimer, defined by two hole-free $CuO_4$ units and two hole-doped $CuO_4$ units, is $4a$, consistent with the estimated size range of *d*-wave Cooper pairs, $4a$–$9a$.[5-8] In each AHD chain (**Fig. 7a**), these protected spin dimers form a beads-on-a-chain structure: the protected spin dimers serve as the beads, while the holes link them together. Thus, AHD chains provide protected channels through which protected spin dimers can move collectively without being destroyed (see Section 6.1 for further discussion).

Thus far, AHDs and AHD chains have been considered along the *a*-direction, where protected spin dimers can move collectively without being destroyed. If analogous chains form along the *b*-direction, the protected spin dimers can likewise move collectively along *b*. By contrast, such chains cannot form along the ($a$+$b$) or (-$a$+$b$) directions, so protected spin dimers cannot propagate along those diagonals. Accordingly, their allowed motion has *d*-wave symmetry in reciprocal-space, provided that current flowing through protected channels along *a* can switch to channels along *b*, and vice versa, as discussed below.

The superconducting gap $\Delta_{sc}$ can be identified with the energy gained when each protected spin dimer in an AHD chain changes from the ↑↓ broken-symmetry state to the singlet state, as discussed in detail in Section 5. Thus, the AHD chain offers a concrete realization of Anderson's conjecture on the cuprates, together with the related consequences (a1-1) and (a1-2).

The above discussion strongly suggests that the protected spin dimers constitute the Cooper pairs in cuprate superconductors. This suggestion implies (e1) that the superconducting current can flow only along the ||*a*- or ||*b*-directions, since the AHD chains run exclusively along those directions. As noted earlier, Wen *et al.*[30] showed that, in the superconducting state, short ||*a*-stripes and short ||*b*-stripes pack together to cover the entire superconducting region. Because these stripes

are composed of AHD chains, a short ||*a*-stripe (||*b*-stripe) corresponds to a "square" segment of short ||*a*-AHD chains (||*b*-AHD chains). For convenience, we refer to these as small ||*a*-segments and small ||*b*-segments. Implication (e1) then leads to two further consequences: (e1-1) the packed arrangement of small ||*a*- and ||*b*-segments enables the superconducting current to flow in all directions within the doped $CuO_2$ layers, and (e1-2) the continuous paths formed by this packing must be of filamentary nature, since the number of holes available to form the ||*a*- and ||*b*-segments is limited, with typical hole densities below 0.3.

### 4.3. Continuous path of packed AHD chain segments and its implications

To prepare for the discussion in the following, we use a simplified representation of the CDW and SDW features of the AHD chain. In **Figs. 8a** and **8b**, hole sites are shown as empty circles and $Cu^{2+}$ spin sites as small dots; up- and down-spins at the $Cu^{2+}$ sites are indicated by cobalt and red arrows, respectively. The spin configurations in **Figs. 8a** and **8b** are equivalent, and both are needed to construct the singlet states of the protected spin dimers discussed (see below). Moreover, the absence of an SDW in neutron diffraction, despite the presence of a CDW (**Fig. 8c**), indicates only that the spin sites have no net spin moment, possibly because of fast spin flips or singlet-state formation. Since the hole sites are electron-deficient relative to the spin sites, we represent the CDW in the AHD chain as a wave whose crests coincide with the spin sites and whose troughs coincide with the hole sites, as shown in **Fig. 8d**.

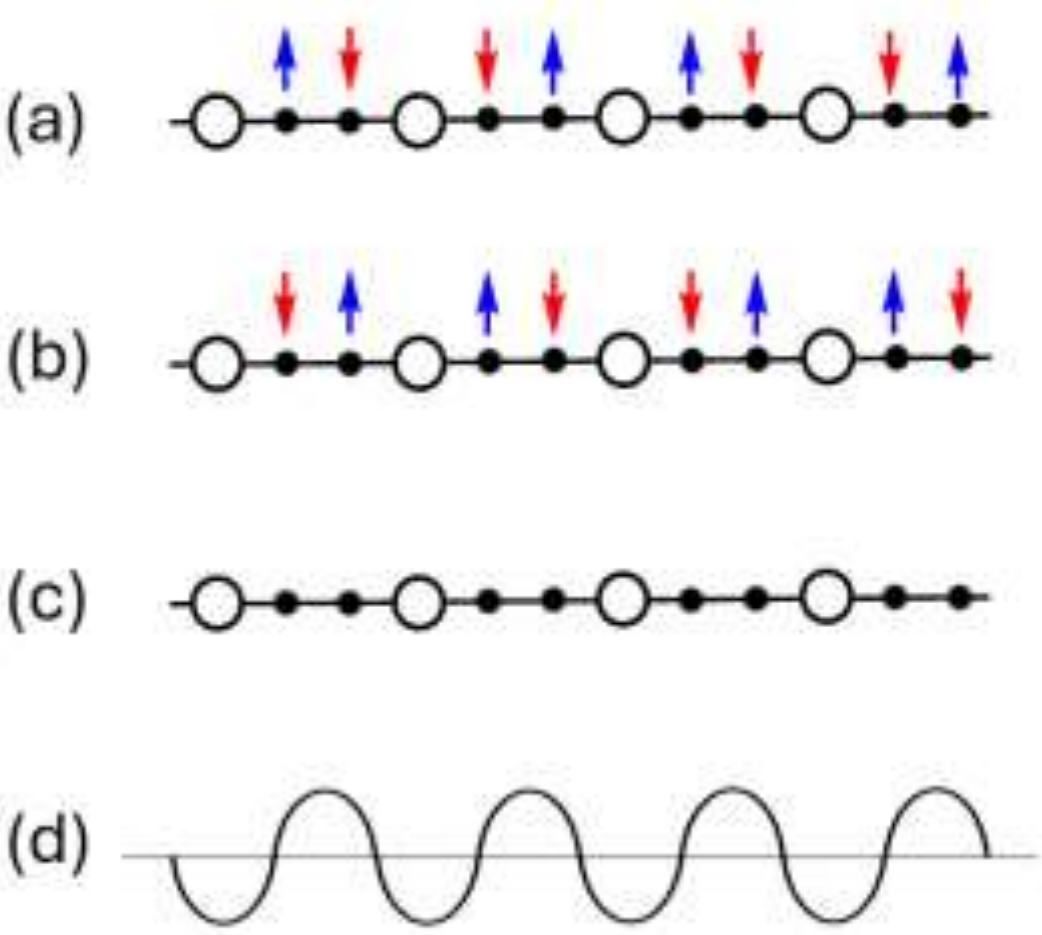

Fig. 8. Simplified notations for describing the features of CDW and SDW that occur in the AHD chains and hence in the stripe phases. In (a) – (c), the empty circles represent the hole-doped $CuO_4$ units, and the dots the hole-free $CuO_4$ units. The up-spins and down-spins at the hole-free sites are indicated by violet and red spins, respectively. The CDW character was represented by a wave with crescents and troughs representing the spin and hole sites, respectively.

The previous section showed that (f1) long-range CDW arises from long AHD chains, whereas short-range CDW corresponds to short AHD chains, and that (f2) AHD chains provide protected channels for superconducting current flow, hereafter referred to as SC-current channels. Wen *et al.'s* observation (d1) therefore implies that (f3) SC current can propagate through the phase dominated by short AHD chains but not through the phase dominated by long AHD chains. We now examine the origin of this contrast.

For this discussion, we recall that $||a$-AHD chains provide only $||a$-SC-current channels, whereas $||b$-AHD chains provide only $||b$-SC-current channels. As shown in **Fig. 9a**, a $||a$-segment consists of short $||a$-AHD chains and therefore contains $||a$-SC-current channels. Similarly, a $||b$-segment consists of short $||b$-AHD chains and contains $||b$-SC-current channels (**Fig. 9b**). If the $||a$- and ||*b*-segments are arranged as in **Fig. 9c**, the ||*a*-SC-current channels in each ||*a*-segment are interrupted by adjacent ||*b*-segments, and conversely; thus, no continuous SC-current channel can form. This simple picture, however, does not account for the possibility that the nodes where two ||*a*-segments and two ||*b*-segments meet can act as Josephson junctions,[61] which allow tunneling of Cooper pairs.

As illustrated in **Fig. 9d**, a slight relative displacement of the four segments can bring several ||*a*-AHD chains and several ||*b*-AHD chains within the coherence distance. Such a node can then function as a Josephson junction, allowing SC current to pass from one segment to another along both the *a*- and *b*-directions. It can also allow the SC current to switch between the *a*- and *b*-directions. If many nodes formed by the assembly of ||*a*- and ||*b*-segments act as internal Josephson junctions, SC current can flow in any direction within the hole-doped $CuO_2$ layers.

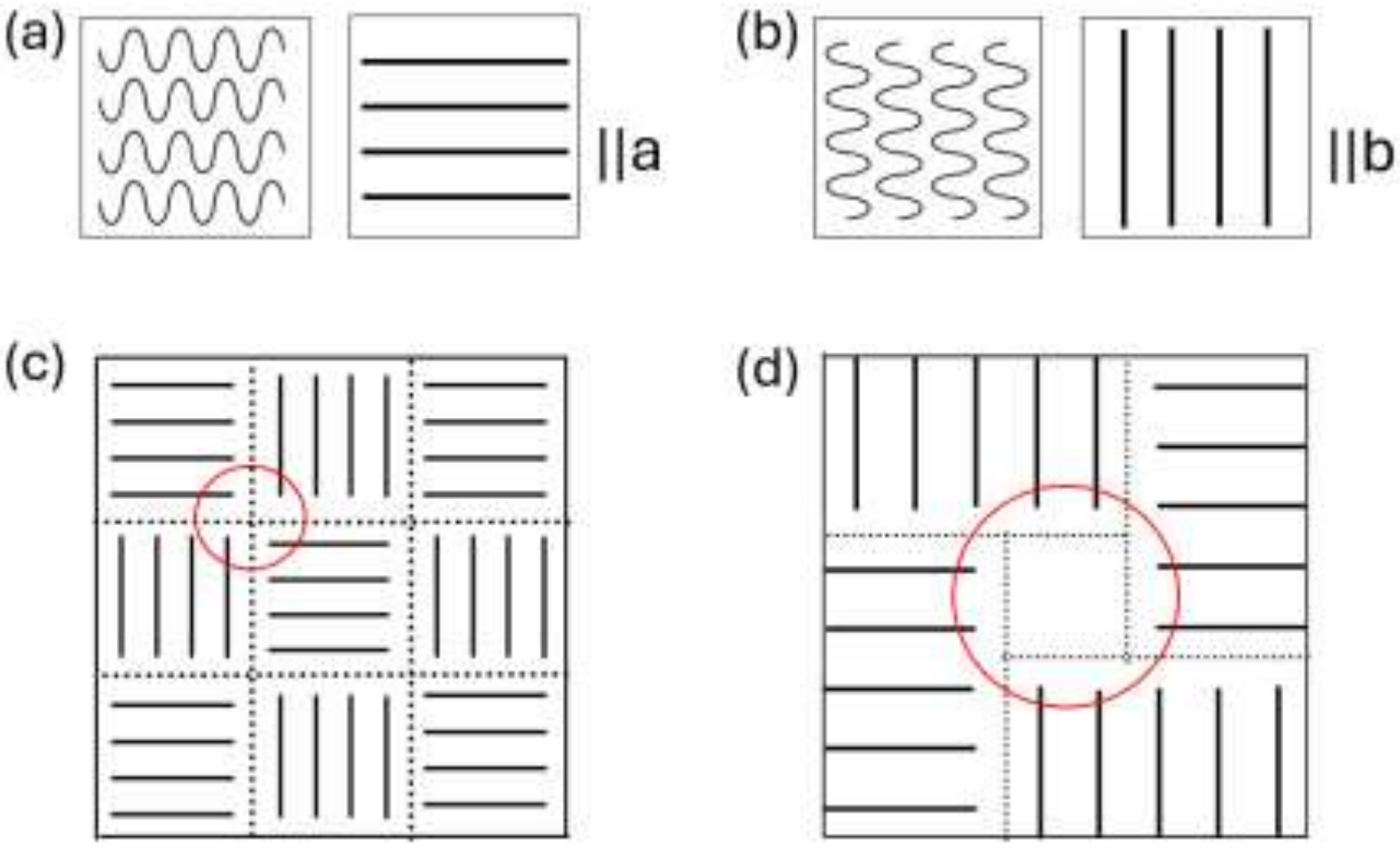

Fig. 9. (a) A simplified notation (right) representing a ||*a*-segment of short AHD chains (left). (b) A simplified notation (right) representing a ||*b*-segment of short AHD chains (left). (c) Assembly of ||*a*- and ||*b*-segments expected in the superconducting phase. (d) Zoomed-view of a node where two ||*a*-segments meet with two ||*b*-segments in a slightly slipped way to form a Josephson junction to allow for Cooper pairs to tunnel through from one segment to another across the node.

The above reasoning clarifies why superconductivity appears in the short-range CDW phase but not in the long-range CDW phase in Wen *et al.*'s high-field X-ray scattering study[30] of LSCO. In the long-range CDW phase, regions composed of ||*a*-AHD chains are separated from those composed of ||*b*-AHD chains. As a result, continuous SC-current channels cannot form, preventing superconductivity in this phase.

Cuprates with hole densities below the minimum value $p_{min}$ are non-superconducting most likely because they lack enough holes to form continuous SC-current channels. Even in superconducting cuprates, the hole density $p$ remains below ~0.3, so the resulting continuous SC-current channels are necessarily filamentary. Cuprates with hole densities above the maximum value $p_{max}$ are also non-superconducting, most likely because the heterogeneity between the hole-doped and hole-free regions in the $CuO_2$ layer becomes too weak to support the formation of AHDs and AHD chains.

An important implication of the above discussion is that, without Josephson tunneling at nodes where two ||*a*-segments meet two ||*b*-segments, SC current flow along AHD chains in the *a*-direction cannot switch to the *b*-direction. Therefore, the experimentally observed *d*-wave symmetry of cuprate Cooper pairs implies that the measured samples contained ||*a*- and ||*b*-segments assembled as in **Fig. 9c**, allowing SC current to flow along both the *a*- and *b*-directions. In samples dominated by long stripes, continuous SC-current channels are absent, so there will be no SC-current flow.

## 5. Nature of the three characteristic energy gaps of cuprate superconductors

Cuprate superconductors exhibit three characteristic excitation energy gaps: the superconducting gap $\Delta_{sc}$, the pseudogap $\Delta_{ps}$, and the gap $\Delta_{nir}$ associated with the center of the nearly flat near-IR absorption peak. In this section, we relate these excitation energies to the spin-exchange interactions that govern the formation, stabilization, and excitation of protected spin dimers in the hole-doped $CuO_2$ layers. To establish this connection, it is crucial to consider the energy spectrum of an AFM spin dimer composed of two $S = 1/2$ ions described by the spin Hamiltonian $H_{spin}$,

$$H_{\text{spin}} = J\, \vec{S}_1 \cdot \vec{S}_2$$

The eigenstates of this spin dimer are the singlet and triplet states $|S\rangle$ and $|T\rangle$, with energies $-3J/4$ and $+J/4$, respectively.[65] The spin-exchange constant $J$ is therefore the energy separation between these states, $J = E_T - E_S$ (**Fig. 10**). Because broken-symmetry states, $|BS\rangle$, are commonly used to describe spin-exchange interactions of magnets, the energy differences involving $|BS\rangle$ states must be related to the singlet-triplet separation $J$, as indicated by the four energy changes labeled (1)–(4) in **Fig. 10**. Energy change (1) corresponds to bringing two initially noninteracting spin sites into an AFM-coupled state $|\uparrow\downarrow\rangle$, giving an energy gain of $J/4$. Energy change (2) transforms the AFM-coupled state $|\uparrow\downarrow\rangle$ into the singlet state $|S\rangle$, producing an additional energy gain of $J/2$. Energy change (3), from the AFM-coupled state $|\uparrow\downarrow\rangle$ to the triplet state $|\uparrow\uparrow\rangle$, requires an energy input of $J/2$. Finally, energy change (4) is the excitation from the singlet state $|S\rangle$ to the triplet state $|T\rangle$.

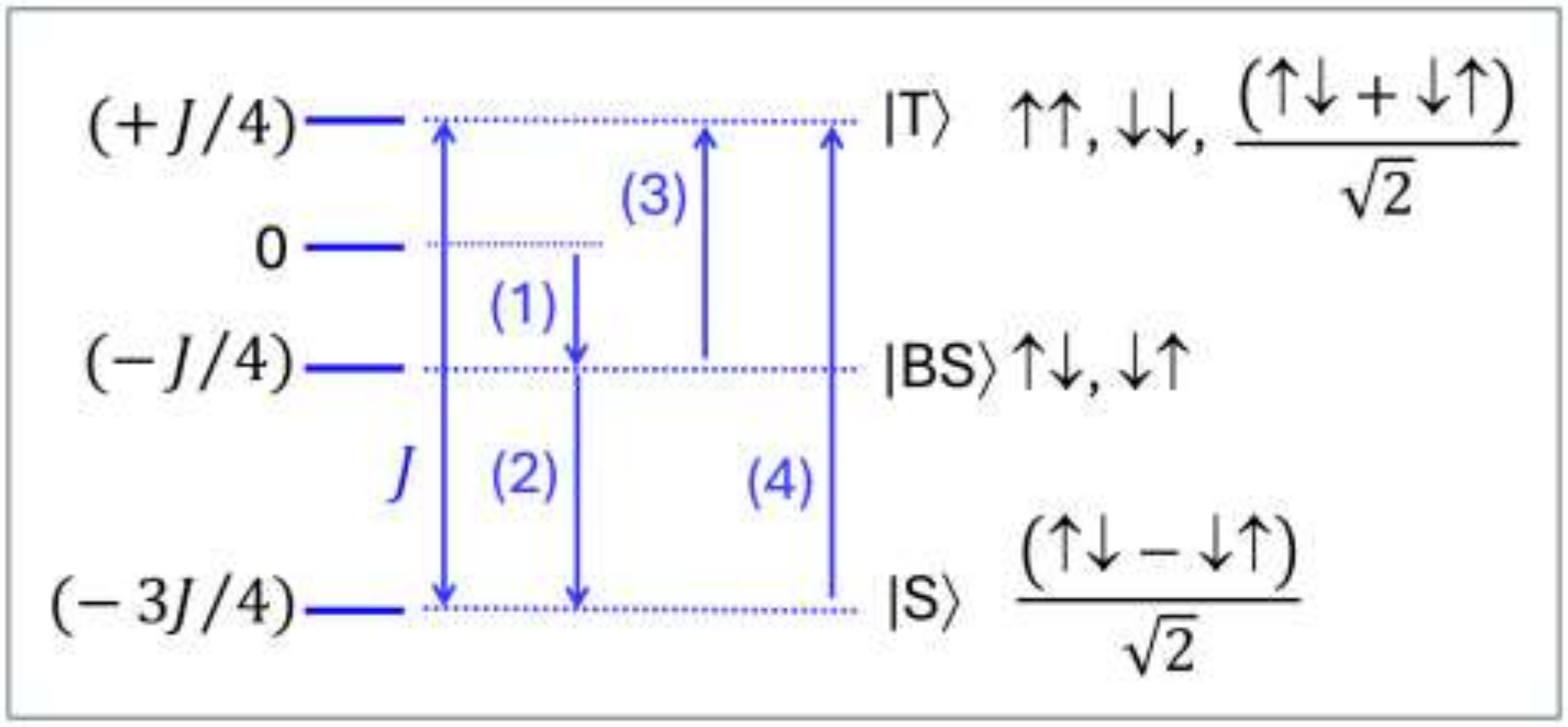

Fig. 10. Energy levels of an antiferromagnetic spin dimer described by $H_{spin}$, showing the singlet and triplet states, the broken-symmetry states, and the four energy differences used to relate BS-state energetics to the spin-exchange constant $J$. The labels S, BS, and T denote the singlet, the broken-symmetry, and the triplet state, respectively.

### 5.1. Superconducting gap $\Delta_{sc}$

So far, each protected spin dimer in an AHD chain has been treated as occupying the broken-symmetry state |BS⟩, represented by ↑↓ or ↓↑. This description accounts for the SDW states observed in the stripe phase. In the superconducting state, however, all protected spin dimers must instead occupy the singlet state |S⟩, that is, the $S = 0$ bosonic state, so that they can act as Cooper pairs. Because the singlet state |S⟩ is lower in energy than the triplet state |T⟩ by $J$ (**Fig. 10**), one might initially identify $\Delta_{sc}$ with $J$. This identification, however, is valid only for a truly isolated spin dimer.

To clarify the discussion in this and the following section, we introduce a simplified notation for the hole and spin arrangements in HMs, AHDs, and AHD chains by slightly modifying the notation used in **Fig. 8a**. If each hole site is denoted by the symbol "+", then the ↑-HM and ↓-

HM can be written as "↑+↑" and "↓+↓", respectively, representing "hole-bound up-spins" and "hole-bound down-spins", respectively. An AHD is then represented as "↑+↑↓+↓" or "↓+↓↑+↑", where a hole-protected spin dimer appears as "+↑↓+" or "+↓↑+". With this notation, **Fig. 11a** represents the |BS⟩ state associated with the CDW and SDW features of an AHD chain, whereas **Figs. 11b** and **11c** represent the corresponding singlet and triplet states. It follows that the lowest-energy excitation available to each spin dimer from the |S⟩ state is the |BS⟩ state, separated by $J/2$; therefore, $\Delta_{sc} = J/2$ (see Section 6.1 for further discussion).

(c) + |T⟩ + |T⟩ + |T⟩ + |T⟩ + |T⟩ +

$J/2$

(a) + ↑↓ + ↓↑ + ↑↓ + ↓↑ + ↑↓ +

$J/2$

(b) + |S⟩ + |S⟩ + |S⟩ + |S⟩ + |S⟩ +

Fig. 11. Three states of the protected spin dimers in the AHD chain: (a) The state in which each spin dimer is initially in the broken-symmetry state. (b) The state in which each spin dimer is in the singlet state. (c) The state in which each spin dimer is in the triplet state.

The spin exchange $J$ discussed above refers specifically to the protected spin dimer $J_{sc}$ and must be distinguished from the spin exchange $J$ of dimers in the hole-free region as illustrated for a chain of four consecutive corner-sharing $CuO_4$ units in **Fig. 12a**.

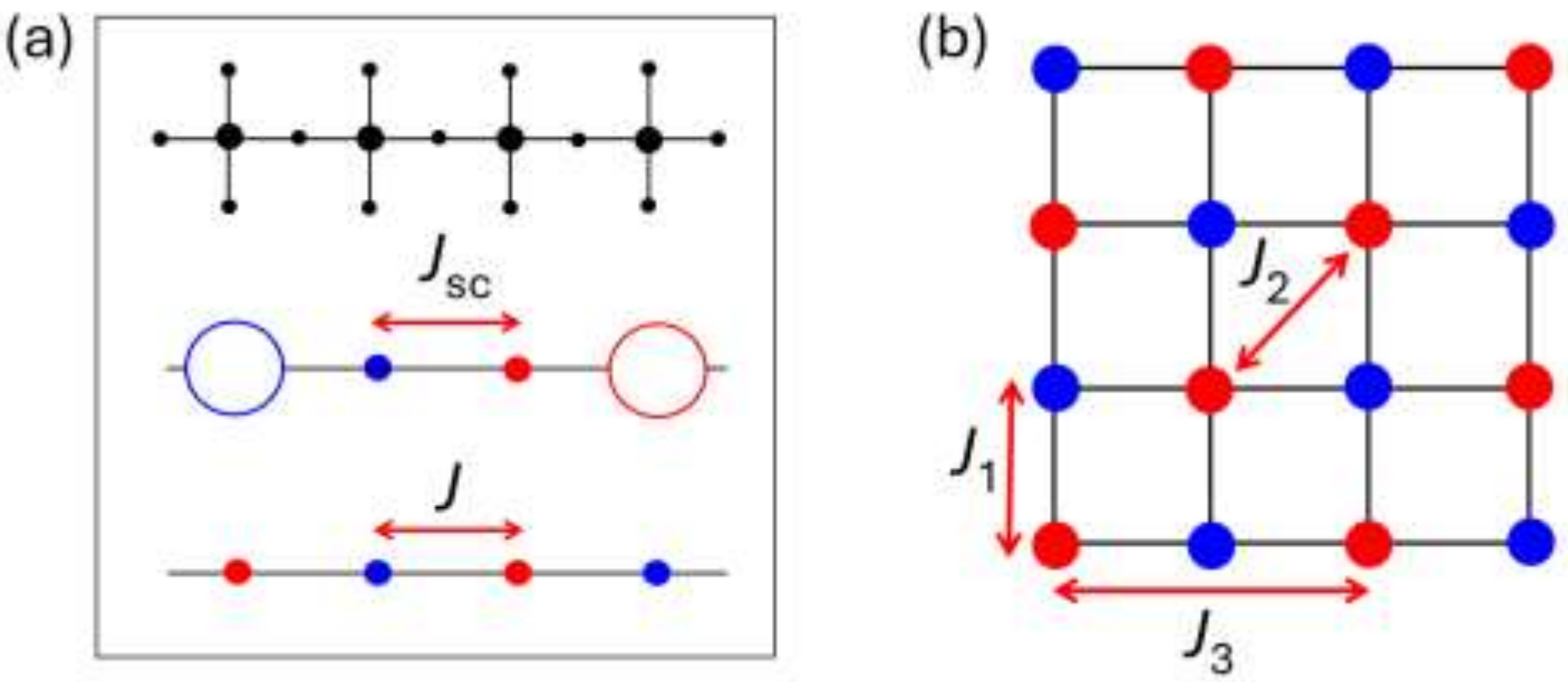

Fig. 12. (a) Distinction between $J_{sc}$ for the hole-protected spin dimer and $J$ for a spin dimer nonadjacent to hole sites in a chain of four corner-sharing $CuO_4$ units. (b) Three spin exchanges, $J_1$–$J_3$, used to describe the magnetic properties of $CuO_2$ layers.

It is important to realize how hole-doped $CuO_4$ units modify the strength of $J_{sc}$ relative to $J$ in the hole-free region. The σ-antibonding arrangement between the Cu $x^2$-$y^2$ orbitals and O 2p orbitals in a chain of four consecutive corner-sharing $CuO_4$ units (**Fig. 12a**, top) is shown schematically in **Fig. 13a**. When the two terminal $CuO_4$ units become hole-doped, their Cu $x^2$-$y^2$ orbitals promote σ-bonding overlaps with adjacent O 2p orbitals, as discussed earlier (**Fig. 1a**). To maximize these σ-overlaps, the $x^2$-$y^2$ orbitals at the hole sites hybridize with the adjacent O 2p orbitals. This enhances σ-bonding of the Cu-O bonds of the hole-doped $CuO_4$ units thereby shortening their bond lengths.[66,67] This stabilization will be important in our later discussion in Section 6.1. This orbital reorganization, in turn, modifies the $x^2$-$y^2$ orbitals of the $Cu^{2+}$ ions that form the protected spin dimer, as indicated in **Fig. 13b**. As a result, the antibonding interaction between the O 2p orbital and the hybridized Cu $x^2$-$y^2$ orbitals along the $J_{sc}$ exchange path is weakened. Consequently, the spin exchange $J_{sc}$ of the protected spin dimer, which is identified with $\Delta_{sc}$, should be smaller than the spin exchange $J$ of spin dimers in the hole-free region.

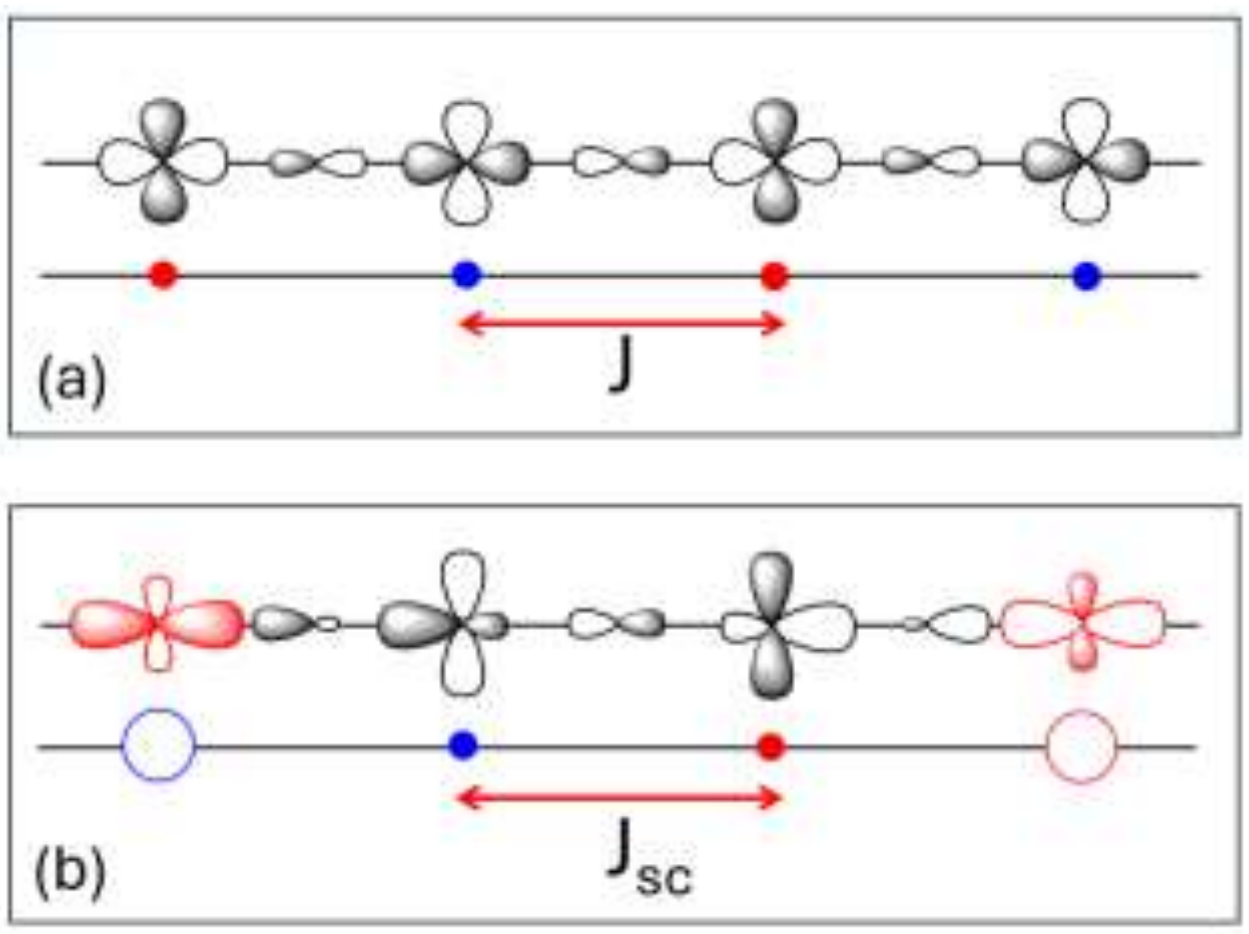

Fig. 13. (a) Cu $x^2$-$y^2$ and O 2p orbitals involved in the spin exchange $J$ between two $Cu^{2+}$ ions nonadjacent to hole sites. (b) Hybridization of the Cu $x^2$-$y^2$ orbitals in the hole-protected spin dimer, which weakens $J_{sc}$. This hybridization arises from the hole-doped $CuO_4$ units maximizing their σ-bonding with adjacent O 2p orbitals.

The magnetic properties of hole-free $CuO_2$ layers are commonly described in terms of three spin-exchange interactions, $J_1$– $J_3$ (**Fig. 11b**). Calculations for various one- to three-layer cuprates showed[78] that $J_1$ lies in the range 93–110 meV, $J_2$ in the range −5 to −11 meV, and $J_3$ is weaker than $J_2$. For simplicity, $J_1$ is denoted as $J$ throughout this work. Experimentally, $J$ for $La_2CuO_4$ has been determined to be 116 meV by Raman scattering[79] and 112 meV by neutron scattering.[80] From these $J$ values, $J/2$ is expected to fall between ~47 and ~58 meV; consequently, the observed $\Delta_{sc}$ should not exceed this range. This expectation is consistent with reported values of 41 meV for $YBa_2Cu_3O_{6.95}$ (YBCO) from neutron scattering,[81] 30 meV for $Bi_2Sr_2CaCu_2O_{8+x}$ (Bi-2212) from angle-resolved photoelectron spectroscopy (ARPES),[82] and 9 and 11 meV for near-optimally doped and underdoped $La_{2-x}Sr_xCuO_4$ (LSCO), respectively.[83]

### 5.2. Nearly flat near-infrared absorption

In the near-IR region, cuprate superconductors exhibit nearly flat absorption with a broad maximum near 600 meV (i.e., $\Delta_{nir} \approx 600$ meV), reflecting low-energy excitations.[84] Hole aggregates, such as AHD chains embedded in hole-free $CuO_4$ regions, are expected to absorb near-IR radiation more strongly than the surrounding hole-free regions because their Cu–O bonds are more polar. For discussion, the spin arrangement in the protected channel of an AHD chain is represented by the bottom diagram labeled G in **Fig. 14**, which denotes the ground state. Consecutive spin flips generate a series of excited states, $E_1$–$E_5$, as labeled in **Fig. 14**, and their effects are summarized in **Fig. 15**. Relative to the ground-state spin configuration in **Fig. 15a**, the first excited state, $E_1$ (**Fig. 15b**), involves three changes from ↑↓ to ↑↑ spin coupling. Each change corresponds to process (3) in **Fig. 10** and requires an energy input of $J/2$. Because two different exchange paths are involved—one along the protected path $J_{sc}$ and the other along the unprotected path $J$—a single spin flip raises the energy by $J + J_{sc}/2$. Two spin flips raise the energy by $2J + J_{sc}$ (**Fig. 15c**). Three spin flips, $E_3$, restore the ↑↓ coupling along one $J_{sc}$ path, giving an excitation energy of $3J + J_{sc}/2$ (**Fig. 15d**). Thus, the excitation energies for n consecutive spin flips are

$$nJ + J_{sc}/2 \qquad \text{for odd n,}$$

$$nJ + J_{sc} \qquad \text{for even n.}$$

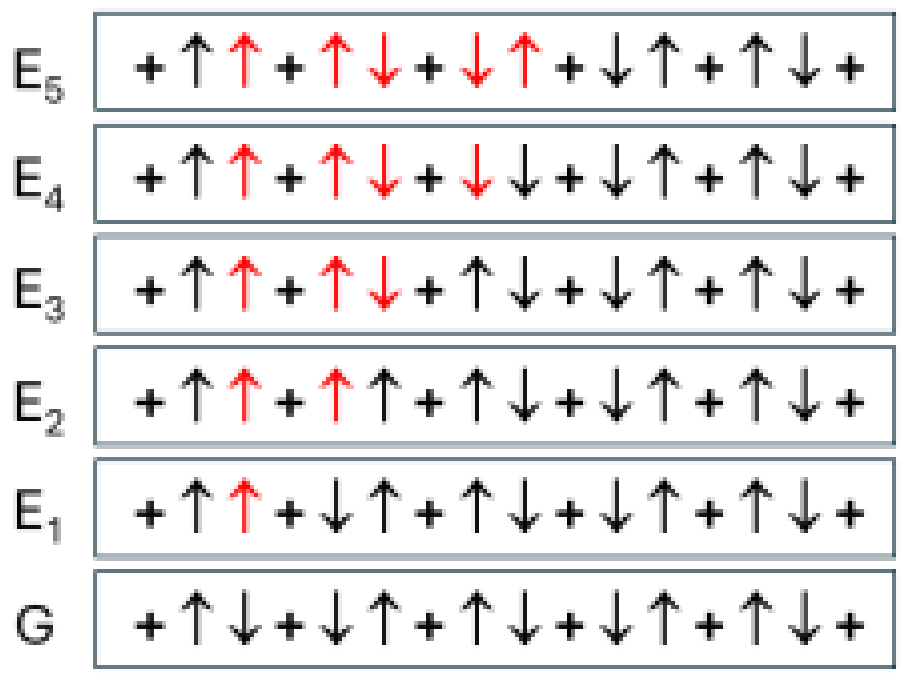

Fig. 14. Simplified representations of the ground state (G) and a series of excited states ($E_1$–$E_5$) generated by consecutive spin flips in the protected channel of an AHD chain.

Consecutive spin flips produce a series of low-lying excited states separated approximately by $J + J_{sc}/2$. This sequence accounts for the nearly flat near-IR absorption observed in hole-doped cuprates because $J \approx 93$–$110$ meV and $J_{sc}$ is smaller than $J$. Taking $J \approx 100$ meV for simplicity, $\Delta_{nir} \approx 600$ meV corresponds to approximately five consecutive spin flips, namely, $\Delta_{nir} \approx 5J + J_{sc}/2$.

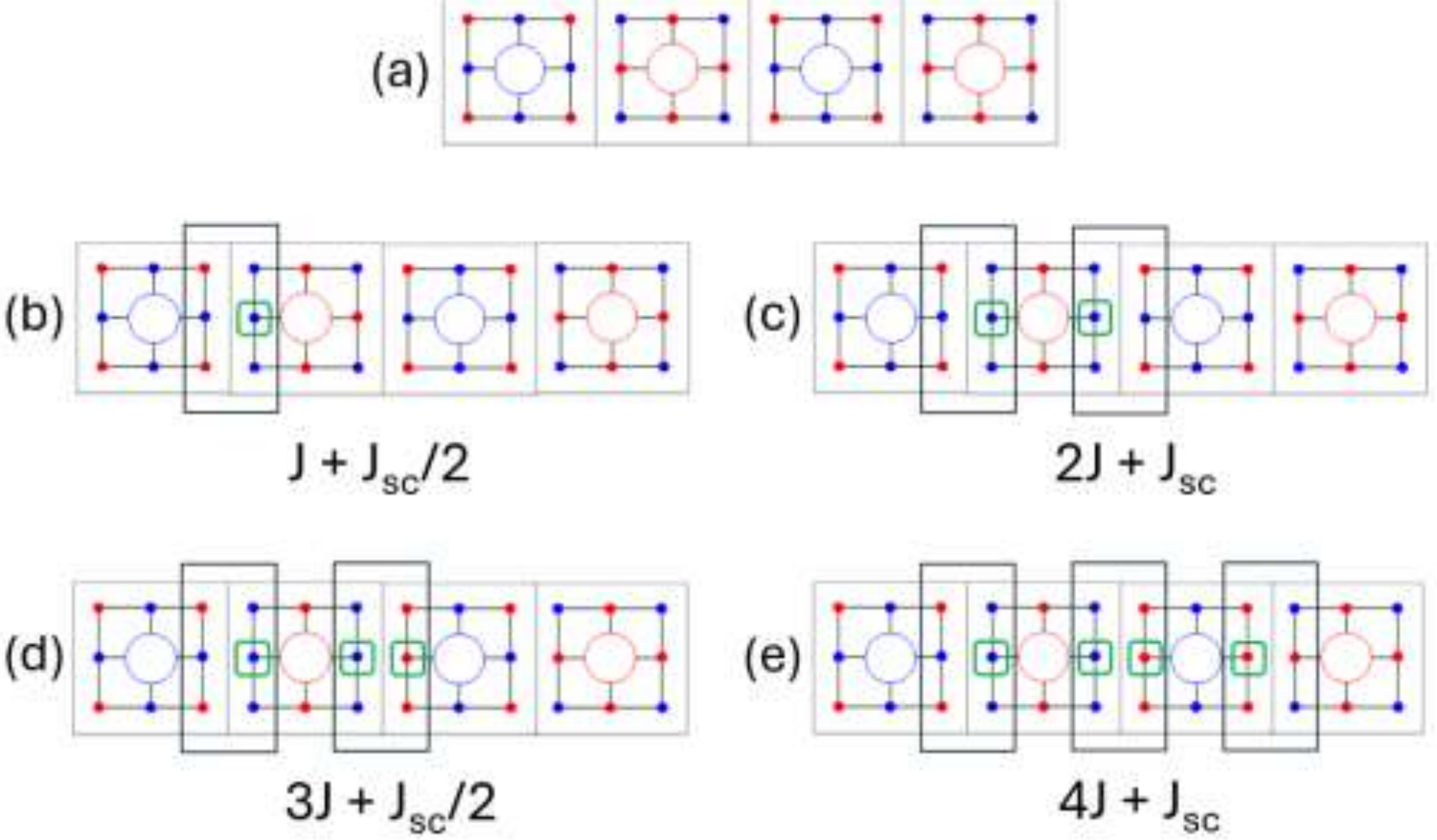

Fig. 15. (a) Spin arrangement of an AHD chain in the ground state. (b)–(e) Consecutive spin flips in the protected channel of spin dimers, producing ↑↓ to ↑↑ spin rearrangements and raising the energies of the excited states $E_1$–$E_4$ relative to the ground state G.

### 5.3. Pseudogap $\Delta_{ps}$

In cuprate superconductors, the pseudogap transition temperature $T^*$ decreases approximately linearly as the hole density $p$ increases.[2,3,34] In Section 4.3, we attributed this behavior to the linear relation between $\Delta T$ and $n_h$ for $k$-stripes formed through the linear growth of hole aggregates from smaller units such as AHDs. This interpretation led to implications (d1) and (d2), and hence to the conclusion that the pseudogap $\Delta_{ps}$ corresponds to the energy gained by AFM coupling between AHD units and is therefore related to the AFM exchange constant $J$. To evaluate this conclusion, consider the AFM coupling of one AHD unit to another, as shown in **Fig. 16**. This coupling corresponds to process (1): it creates two ↑↓ magnetic bonds governed by $J$ and one ↑↓ magnetic bond governed by $J_{sc}$, giving an energy stabilization of $J/2 + J_{sc}/4$. Thus, $\Delta_{ps} = J/2 + J_{sc}/4$. With $J \approx 93–110$ meV and $J_{sc} \approx J/4$, this expression gives $\Delta_{ps} \approx 70–83$ meV. This estimate agrees with experimental values of up to ~70 meV for Bi-2212[85] and ~50–70 meV for YBCO, LSCO, and $Y_{0.8}Ca_{0.2}Ba_2Cu_3O_{7-\delta}$.[84]

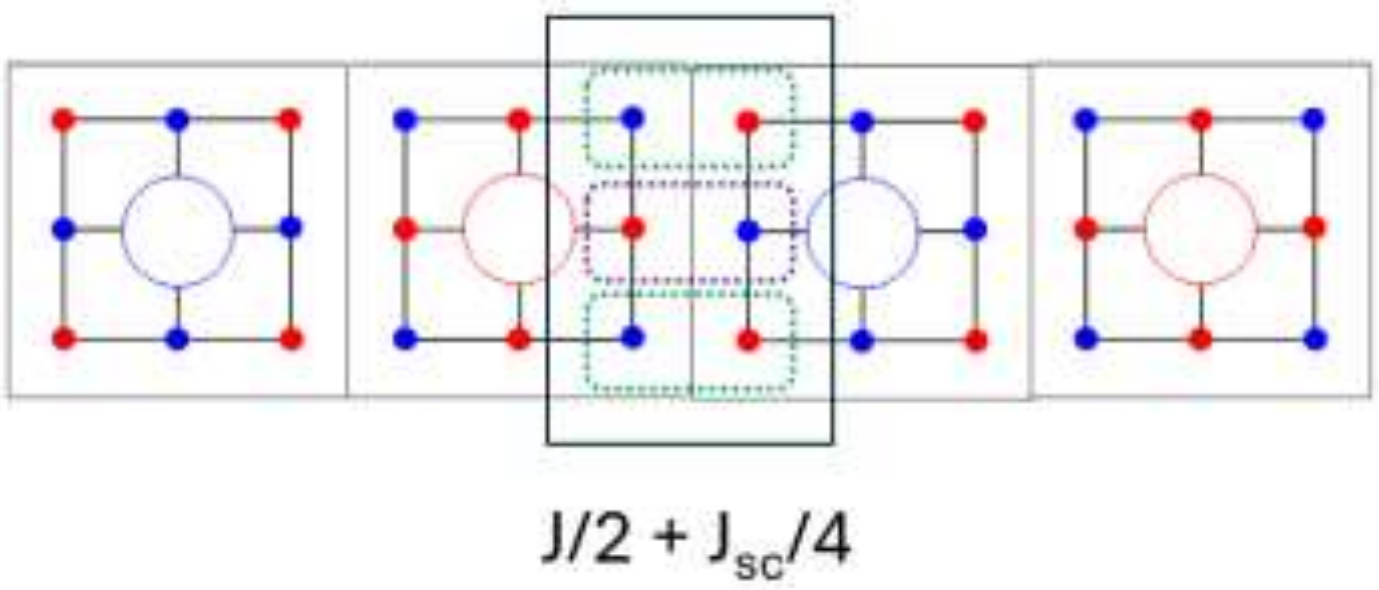

Fig. 16. Energy stabilization produced when two AHD units combine into a larger aggregate through AFM coupling. This coupling creates three new antiferromagnetic spin-exchange paths.

According to our discussions so far, the three characteristic energy gaps are related to the spin exchanges of hole-doped $CuO_2$ layers as follows:

$$\Delta_{sc} \approx J_{sc}/2$$

$\Delta_{ps} \approx J/2 + J_{sc}/4$

$\Delta_{nir} \approx 5J + J_{sc}/2$

That is, they are all linked to spin-exchange interactions in the $CuO_2$ layers, as modified by the formation of hole aggregates. The resulting energy hierarchy is $\Delta_{sc} < \Delta_{ps} << \Delta_{nir}$. These findings support implication (a2) derived from Anderson's conjecture.[9]

**6. Discussion**

Our search for the probable real-space shape and size of *d*-wave Cooper pairs in high-$T_c$ cuprate superconductors was guided by two key ideas: Anderson's conjecture on spin-exchange-coupled Cooper pairs and the observation of electronic phase separation in doped $CuO_2$ layers. We analyzed the segregation of hole aggregates in hole-doped $CuO_2$ layers based on their segregation pressure indices to find that each AHD contains an antiferromagnetic spin dimer protected by two holes, and that each AHD chain has a beads-on-a-chain structure, with the protected spin dimers as the beads and the holes linking them together. Wen et al.'s observation of coexisting superconductivity and short-range CDW order in LSCO was crucial to our proposal that these protected spin dimers are *d*-wave Cooper pairs and that the AHD chains provide protected channels through which the spin dimers can move collectively without being destroyed. Because these channels allow superconducting current to flow only along the $||a$- or $||b$-direction, it was necessary to identify a mechanism that allows SC current to switch from the $||a$- to the $||b$-channels, and vice versa. The key clue to this mechanism was also provided by Wen et al.'s finding that, in regions where superconductivity coexists with short-range CDW order, square segments of short $||a$- and $||b$-stripes pack together orthogonally, covering the superconducting region and forming nodes at every segment corner. This finding led us to propose that these nodes act as Josephson junctions, providing the mechanism for superconducting current to flow in all directions within hole-doped $CuO_2$ layers. In the following we examine several important remaining issues.

### 6.1. Role of the singlet state in a protected spin dimer

In the ground state of the spin arrangement of an AHD chain, which describes the CDW/SDW features of the stripes, each protected spin dimer couples antiferromagnetically to the spins in its surrounding protective shell. Its singlet state, |S⟩, is the linear combination of the two broken-symmetry states |↑↓⟩ and |↓↑⟩:

$$|\mathrm{S}\rangle = \frac{|\uparrow\downarrow\rangle - |\downarrow\uparrow\rangle}{\sqrt{2}}.$$

As illustrated in **Fig. 17a**, when the spin dimer is represented by the |↑↓⟩ state, it forms four ↑↓ exchange couplings with the spins of the protective shell. When represented by the |↓↑⟩ state, it instead forms four ↑↑ exchange couplings. In the singlet state |S⟩, these contributions cancel, leaving no net spin-exchange interaction between the dimer and the protective shell. The resulting chain of bosonic spin dimers (i.e., $S = 0$ entities) can therefore move as Cooper pairs through the channel bounded by the two fermion chains (i.e., chains of $S = 1/2$ entities). This explanation raises an important question. The broken-symmetry state |BS⟩ of a protected spin dimer, shown in the left part of **Fig. 17a**, is more stable than the singlet state |S⟩ with respect to coupling to the protective shell: the |BS⟩ state produces four ↑↓ exchange couplings between the protected spin dimer and the shell spins, whereas these couplings vanish in the singlet state. The corresponding energy difference is $J$, because it is of type (1) in **Fig. 10**. Since the singlet state is lower in energy than the broken-symmetry state by $J_{sc}/2$ within the protected dimer, what additional factor provides an energy gain greater than $J - J_{sc}/2$ and thereby favors singlet formation? Most likely, the

hybridization associated with σ-bonding between the Cu $x^2-y^2$ orbital at a hole site and the adjacent O 2p orbital is enhanced when spin exchange between the bosonic chain and its neighboring fermionic chains is suppressed.

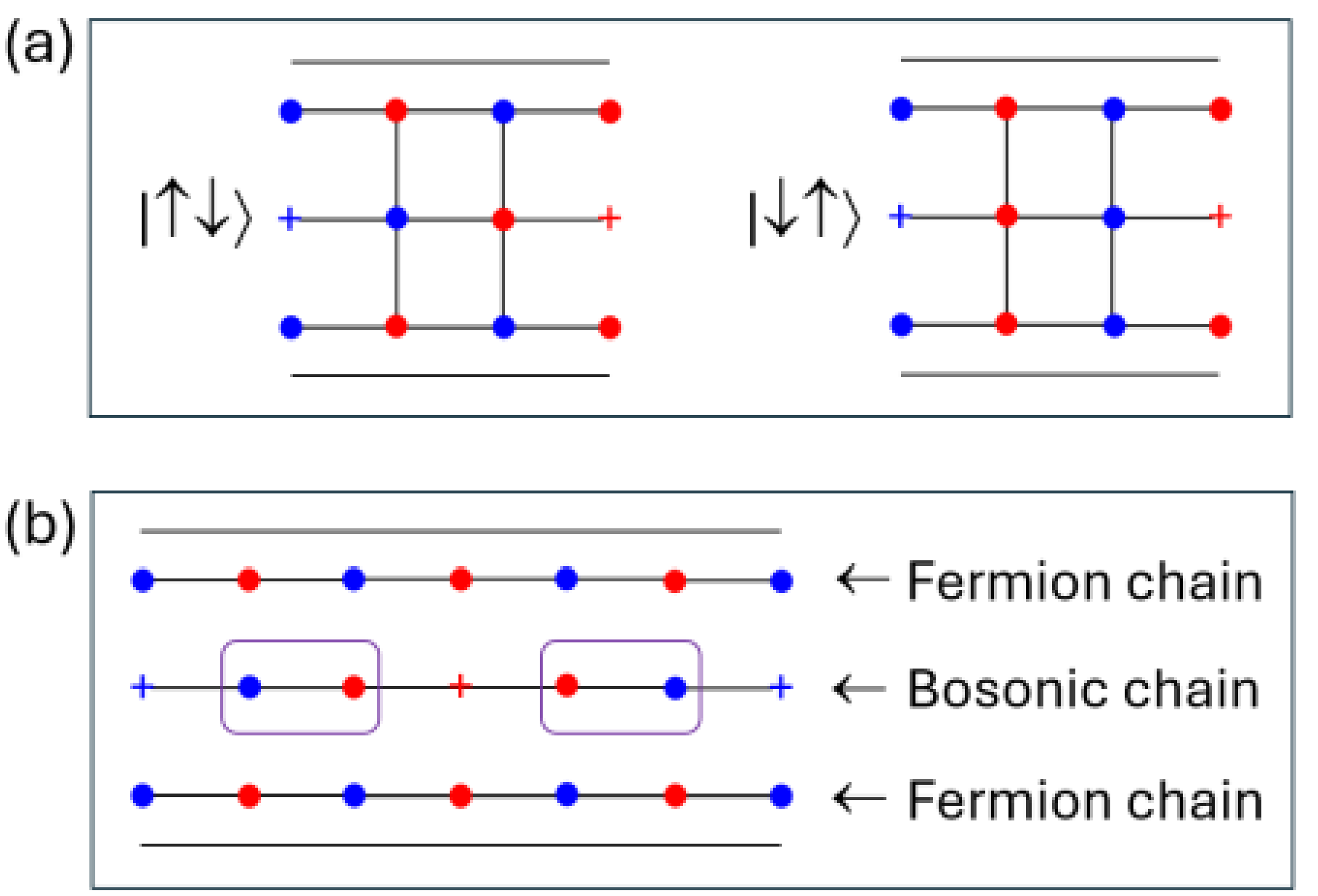

Fig. 17. (a) Spin-exchange interactions between a protected spin dimer and the spins of its protective shell when the dimer is represented by the broken-symmetry state |↑↓⟩ (left) or |↓↑⟩ (right). (b) A bosonic chain of singlet spin dimers that is free from spin-exchange interactions with the surrounding fermionic chains in an AHD chain. The symbols "+" represent the hole sites, and encircled spin dimers indicate that they are in singlet state.

### 6.2. CDW and SDW periods

As discussed in Section 4.1, the commensurate CDW and SDW periods of an AHD chain are $3a$ and $6a$, respectively (**Fig. 7a**). As shown in **Figs. 7b** and **7c**, an AHD can be modified by inserting linear spin trimers between the ↑-HM and ↓-HM units. When such a modified AHD serves as the repeat unit of a chain, the CDW and SDW periods deviate from $3a$ and $6a$, respectively. For example, the modified AHD unit in **Fig. 7b** produces a chain period of $4a$. If an unmodified AHD unit combines with the modified AHD unit in **Fig. 7c** to form a new repeat unit, the resulting chain has a period of $8a$. Such modified AHD chains are more likely to occur at low hole concentrations. Indeed, stripes with these periods are observed in cuprates with low hole concentrations. Periods of $4a$ and $8a$ have previously been explained in terms of a pair-density wave (PDW).[86-88]

In our proposal, AHD chains are the only protected regions of hole-doped $CuO_2$ layers in which *d*-wave Cooper pairs form when the magnetic state of each protected spin dimer changes from the broken-symmetry state $|\uparrow\downarrow\rangle$ to the singlet state $|S\rangle$. In discussing the superconducting state, we implicitly assumed that, below $T_c$, all protected spin dimers become *d*-wave Cooper pairs. In an external magnetic field, however, some Cooper pairs may be broken; these broken pairs are described not by the singlet state $|S\rangle$ but by the broken-symmetry state $|\uparrow\downarrow\rangle$.

### 6.3. Strange-metallic state

Let us consider the magnetic states of AHD chains above $T_c$, where all protected spin dimers occupy broken-symmetry states. The ground state of each AHD chain above $T_c$, shown in **Fig. 11a**, corresponds to antiferromagnetic coupling between the $\uparrow$-HM and $\downarrow$-HM units. Above $T_c$, each AHD chain can therefore undergo spin fluctuations driven either by local spin flips, as discussed in Section 5.2, or by local excitation of individual spin dimers from the $|\uparrow\downarrow\rangle$ state to the $|T\rangle$ state. The AHD chains thus form antiferromagnetic puddles that remain spatially fixed within the hole-doped $CuO_2$ layers, while their spin fluctuations interact with the conducting electrons. Such interactions have been shown[89] to produce the linear temperature dependence of the resistivity $\rho$, characteristic of the strange-metallic state. This reasoning applies to any hole aggregate formed by the aggregation of AHDs in a hole-doped $CuO_2$ layer, regardless of whether it participates in superconductivity. This broader applicability may explain why the strange-metallic region widens with increasing temperature in the *T-p* phase diagram.[2,3]

## 7. Concluding remarks

We examined the chemical aspects of high-$T_c$ superconductivity after identifying the probable real-space shape and size of its *d*-wave Cooper pairs. This work led to the following conclusions:

(a) The *d*-wave Cooper pairs are antiferromagnetic spin dimers protected by two holes in antiferromagnetic hole dimers, or AHDs.

(b) AHD chains provide protected channels through which spin dimers in singlet state can move collectively without being destroyed; these channels carry the superconducting current.

(c) For cuprate superconductivity to occur in hole-doped $CuO_2$ layers, square segments composed of short AHD chains aligned along $||a$ and $||b$ must pack orthogonally to form continuous paths. The nodes between the $||a$ and $||b$ segments then act as internal Josephson junctions.

(d) The pseudogap state is associated with the linear growth of stripe phases, and the linear $T^*$–$p$ relationship reflects that the reorganization energy per hole required for a smaller linear aggregate to become a larger one is constant.

(e) We identified the nature of the superconducting gap, the pseudogap, and the near-IR absorption gap and found that they are related to the nearest-neighbor spin exchanges $J$ and $J_{sc}$ as follows: $\Delta_{sc} \approx J_{sc}/2$, $\Delta_{ps} \approx J/2 + J_{sc}/4$, and $\Delta_{nir} \approx 5J + J_{sc}/2$.

(f) The linear resistivity $\rho \propto T$ observed in the strange metallic state above $T_c$ is caused by hole aggregates made up of AHDs, which act as antiferromagnetic puddles. Their spin fluctuations, driven either by local spin flips or by local excitations of individual spin dimers from the $|\uparrow\downarrow\rangle$ state to the $|T\rangle$ state, can interact with conducting electrons.

(g) Our work shows that Anderson's conjecture is correct.

In summary, our work treated the structural, phase-segregation, charge/spin-order, and superconductivity aspects of cuprate superconductors as interconnected elements of a single real-space picture. Beyond addressing the puzzling features of high-$T_c$ cuprate superconductors discussed above, our analysis of the AHD chain yields an interesting conceptual picture (**Fig. 17**). An AHD chain comprises three parallel chains: two outer AFM chains formed by spins in the protective shells of the HM units and a central chain of holes and protected spin dimers. Because the outer chains consist of $S = 1/2$ ions, they are fermionic. When each protected spin dimer adopts the broken-symmetry state, $|\uparrow\downarrow\rangle$ or $|\downarrow\uparrow\rangle$, the central "hole/spin-dimer" chain is represented by $(+\uparrow\downarrow+\downarrow\uparrow)_\infty$ and is also fermionic. When each protected spin dimer instead adopts the singlet state, the central chain, represented by $(+|S\rangle+|S\rangle)_\infty$, becomes bosonic because each singlet spin dimer has $S = 0$. Consequently, the bosonic chain is decoupled from spin-exchange interactions with the two flanking fermionic chains, allowing its bosonic entities—the $d$-wave Cooper pairs—to move unhindered under an applied voltage gradient.

## ■ AUTHOR INFORMATION

**Corresponding Author**

Myung-Hwan Whangbo − *Department of Chemistry, North Carolina State University, Raleigh, North Carolina 27695-8204, USA*; orcid.org/0000-0002-2220-1124; Email: mike_whangbo@ncsu.edu

**Author**

Reinhard K. Kremer − Max Planck Institute for Solid State Research, Heisenbergstrasse 1, D-70569 Stuttgart, Germany; orcid.org/0000-0001-9062-2361; Email: rekre@fkf.mpg.de

## ■ Conflicts of interest

There are no conflicts to declare

## References

(1) Bednorz, J. G.; Müller, K. A. Possible high $T_C$ superconductivity in the Ba-La-Cu-O system. *Z. Physik B Condens. Matter* **1986**, *64*, 189–193.

(2) Keimer, B.; Kivelson, S. A.; Norman, M. R.; Uchida, S.; Zaanen, J. From quantum matter to high-temperature superconductivity in copper oxides. *Nature* **2015**, *518*, 179–186.

(3) Hiroi, Z. Introduction to high-temperature superconductivity for solid state chemists. *Progress in Solid State Chemistry* **2026**, *83*, 100574.

(4) Watson, J. D. *The Double Helix: A Personal Account of the Discovery of the Structure of DNA*; Atheneum Press: USA, 1968.

(5) Carillo, F.; Papari, G.; Stornaiuolo, D.; Born, D.; Montemurro, D.; Pingue, P.; Beltram, F.; Tafuri, F. Little-Parks effect in single nanoscale $YBa_2Cu_3O_{6+x}$ rings. *Phys. Rev. B* **2010**, *81*, 054505.

(6) Petrenko, E. V.; Omelchenko, L. V.; Kolesnichenko, Y. A.; Shytov, N. V.; Rogacki, K.; Sergeyev, D. M.; Solovjov, A. L. Study of fluctuation conductivity in $YBa_2Cu_3O_{7-\delta}$ films in strong magnetic fields. *Low Temp. Phys.* **2021**, *47*, 1050–1057.

(7) Hwang, J. Superconducting coherence length of hole-doped cuprates obtained from electron–boson spectral density function. *Sci. Rep.* **2021**, *11*, 11668.

(8) Chen, G.; Zhang, Y.; Xi, G.; Shen, J.; Wu, J. Intertwined nematic and d-wave superconductive orders in optimally doped $La_{1.84}Sr_{0.16}CuO_4$ thin films. *National Sci. Rev.* **2026**, *13*, nwag309.

(9) Anderson, P. W. Last Words on the Cuprates. 2016; https://arxiv.org/ abs/1612.03919.

(10) Hizhnyakov, V.; Sigmund, E. High-$T_c$ superconductivity induced by ferromagnetic clustering. *Physica C: Supercond.* **1988**, *156*, 655–666.

(11) Hizhnyakov, V.; Kristoffel, N.; Sigmund, E. On the percolation induced conductivity in high-$T_c$ superconducting materials. *Physica C: Supercond.* **1989**, *160*, 119–123.

(12) Kremer, R. K.; Sigmund, E.; Hizhnyakov, V.; Hentsch, F.; Simon, A.; Müller, K. A.; Mehring, M. Percolative phase separation in $La_2CuO_{4+\delta}$ and $La_{2-x}Sr_xCuO_4$. *Z. Physik B Condens. Matter* **1992**, *86*, 319–324.

(13) Müller, K. A.; Benedek, G. *Proceedings of the Workshop on Phase Separation in Cuprate Superconductors*; World Scientific Singapore New Jersey London Honkong, 1993.

(14) Emery, V.; Kivelson, S. Frustrated electronic phase separation and high-temperature superconductors. *Physica C: Supercond.* **1993**, *209*, 597–621.

(15) Sigmund, E.; Müller, K. A. *Phase separation in cuprate superconductors: Proceedings of the second international workshop. September 4-10, 1993, Cottbus, Germany (2012)*; Springer-Verlag Berlin Heidelberg New York, 1994.

(16) Kivelson, S. A.; Emery, V. J. *Electronic Phase Separation and High Temperature Superconductors*; Report BNL-61338, 1994.

(17) Sigmund, E.; Hizhnyakov, V.; Kremer, R. K.; Simon, A. On the existence of percolative phase separation in high-$T_c$ cuprates. *Z. Physik B Condens.Matter* **1994**, *94*, 17–20.

(18) Kivelson, S.; Emery, V. Topological doping of correlated insulators. *Synthetic Metals* **1996**, *80*, 151–158.

(19) Pan, S. H.; O'Neal, J. P.; Badzey, R. L.; Chamon, C.; Ding, H.; Engelbrecht, J. R.; Wang, Z.; Eisaki, H.; Uchida, S.; Gupta, A. K.; Ng, K.-W.; Hudson, E. W.; Lang, K. M.; Davis, J. C. Microscopic electronic inhomogeneity in the high-$T_c$ superconductor $Bi_2Sr_2CaCu_2O_{8+x}$. *Nature* **2001**, *413*, 282–285.

(20) de Mello, E. V. L.; Caixeiro, E. S. Effects of phase separation in the cuprate superconductors.

*Phys. Rev. B* **2004**, *70*, 224517.

(21) McElroy, K.; Lee, J.; Slezak, J. A.; Lee, D.-H.; Eisaki, H.; Uchida, S.; Davis, J. C. Atomic-Scale Sources and Mechanism of Nanoscale Electronic Disorder in $Bi_2Sr_2CaCu_2O_{8+\delta}$. *Science* **2005**, *309*, 1048–1052.

(22) Deutscher, G.; de Gennes, P.-G. A spatial interpretation of emerging superconductivity in lightly doped cuprates. *Comptes Rendus. Physique* **2007**, *8*, 937–941.

(23) Pasupathy, A. N.; Pushp, A.; Gomes, K. K.; Parker, C. V.; Wen, J.; Xu, Z.; Gu, G.; Ono, S.; Ando, Y.; Yazdani, A. Electronic Origin of the Inhomogeneous Pairing Interaction in the High-$T_C$ Superconductor $Bi_2Sr_2CaCu_2O_{8+x}$. *Science* **2008**, *320*, 196–201.

(24) Haase, J.; Rybicki, D.; Slichter, C. P.; Greven, M.; Yu, G.; Li, Y.; Zhao, X. Two-Component uniform spin susceptibility of superconducting $HgBa_2CuO_{4+\delta}$ single crystals measured using $^{63}$Cu and $^{199}$Hg nuclear magnetic resonance. *Phys. Rev. B* **2012**, *85*, 104517.

(25) Storey, J. G.; Tallon, J. L. Two-component electron fluid in underdoped high-$T_c$ cuprate superconductors. *Europhys. Lett.* **2012**, *98*, 17011.

(26) Bill, A.; Hizhnyakov, V.; Kremer, R. K.; Seibold, G.; Shelkan, A.; Sherman, A. Phase Separation and Pairing Fluctuations in Oxide Materials. *Condens. Matter* **2020**, *5(4)* 65.

(27) Hizhnyakov, V.; Seibold, G. Nanoscale phase separation in cuprate superconductors. *Physica C: Supercond. Appl.* **2023**, *612*, 1354309.

(28) Ye, S.; Zou, C.; Yan, H.; Ji, Y.; Xu, M.; Dong, Z.; Chen, Y.; Zhou, X.; Wang, Y. The emergence of global phase coherence from local pairing in underdoped cuprates. *Nature Phys.* **2023**, *19*, 1301–1307.

(29) Sinha, A.; Wietek, A. Forestalled phase separation as the precursor to stripe order. *Nature Commun.* **2025**, *16*, 10807.

(30) Wen, J.-J.; He, W.; Jang, H.; Nojiri, H.; Matsuzawa, S.; Song, S.; Collet, M.; Zhu, D.; Liu, Y.-J.; Fujita, M.; Jiang, J. M.; Rotundu, C. R.; Kao, C,-C.; Jiang, H.-C.; Lee, J.-S.; Lee, Y. S. Enhanced charge density wave with mobile superconducting vortices in $La_{1.885}Sr_{0.115}CuO_4$. *Nature Commun.* **2023**, *14*, 733.

(31) Warren Jr., W.; Walstedt, R.; Brennert, G.; Cava, R.; Tycko, R.; Bell, R.; Dabbagh, G. Cu spin dynamics and superconducting precursor effects in planes above $T_C$ in $YBa_2Cu_3O_{6.7}$. *Phys. Rev. Lett.* **1989**, *62*, 1193–1196.

(32) Walstedt, R. E.; Warren, W. W. Nuclear Resonance Properties of $YBa_2Cu_3O_{6-x}$ Superconductors. *Science* **1990**, *248*, 1082–1087.

(33) Alloul, H.; Mendels, P.; Casalta, H.; Marucco, J. F.; Arabski, J. Correlations between magnetic and superconducting properties of Zn-substituted $YBa_2Cu_3O_{6+x}$. *Phys. Rev. Lett.* **1991**, *67*, 3140–3143.

(34) Batlogg, B.; Hwang, H.; Takagi, H.; Cava, R.; Kao, H.; Kwo, J. Normal state phase diagram of $(La,Sr)_2CuO_4$ from charge and spin dynamics. *Physica C: Supercond.* **1994**, *235-240*, 130–133.

(35) Ding, H.; Yokoya, T.; Campuzano, J. C.; Takahashi, T.; Randeria, M.; Norman, M. R.; Mochiku, T.; Kadowaki, K.; Giapintzakis, J. Spectroscopic evidence for a pseudogap in the normal state of underdoped high-Tc superconductors. *Nature* **1996**, *382*, 51–54.

(36) Mesot, J.; Furrer, A. The crystal field in rare earth based high-temperature supeconductors. *J. Supercond.* **1997**, *10*, 623–643.

(37) Renner, C.; Revaz, B.; Genoud, J.-Y.; Kadowaki, K.; Fischer, O. Pseudogap Precursor of the Superconducting Gap in Under- and Overdoped $Bi_2Sr_2CaCu_2O_{8+\delta}$. *Phys. Rev. Lett.* **1998**, *80*, 149–152.

(38) Randeria, M.; Trivedi, N. Pairing correlations above $T_C$ and pseudogaps in underdoped cuprates. *J. Phys. Chem. Solids* **1998**, *59*, 1754–1758.

(39) Sadovskii, M. V. Pseudogap in high-temperature superconductors. *Phys. Usp.* **2001**, *44*, 515–539.

(40) Timusk, T. The mysterious pseudogap in high temperature superconductors: an infrared view. *Solid State Commun.* **2003**, *127*, 337–348.

(41) Kresin, V. Z.; Ovchinnikov, Y. N.; Wolf, S. A. Inhomogeneous superconductivity and the “pseudogap” state of novel superconductors. *Phys. Rep.* **2006**, *431*, 231–259.

(42) Kawasaki, S.; Lin, C.; Kuhns, P. L.; Reyes, A. P.; Zheng, G.-q. Carrier-Concentration Dependence of the Pseudogap Ground State of Superconducting $Bi_2Sr_{2-x}La_xCuO_{6+\delta}$ Revealed by $^{63,65}$Cu-Nuclear Magnetic Resonance in Very High Magnetic Fields. *Phys. Rev. Lett.* **2010**, *105*, 137002.

(43) Moon, S. J.; Lee, Y. S.; Schafgans, A. A.; Chubukov, A. V.; Kasahara, S.; Shibauchi, T.; Terashima, T.; Matsuda, Y.; Tanatar, M. A.; Prozorov, R.; Thaler, A.; Canfield, P. C.; Bud’ko, S. L.; Sefat, A. S.; Mandrus, D.; Segawa, K.; Ando, Y.; Basov, D. N. Infrared pseudogap in cuprate and pnictide high-temperature superconductors. *Phys. Rev. B* **2014**, *90*, 014503.

(44) Varma, C. M. Pseudogap in cuprates in the loop-current ordered state. *J. Phys.: Condens.Matter* **2014**, *26*, 505701.

(45) Solovjov, A. L.; Vovk, R. V.; Rogacki, K. Unusual behavior of pseudogap in high-temperature superconductors under the influence of external factors. Part I (Review Article). *Low Temp. Phys.* **2025**, *51*, 1061–1079.

(46) Nikolaevsky, M.; Samanta, A.; Trivedi, N.; Frydman, A. The nature of the “pseudogap” in the insulating phase of highly disordered superconductors. 2026;

https://arxiv.org/abs/2608.12508.

(47) Wen, C.; Hou, Z.; Akbari, A.; Chen, K.; Hong, W.; Yang, H.; Eremin, I.; Li, Y.; Wen, H.-H. Unprecedentedly large gap in $HgBa_2Ca_2Cu_3O_{8+\delta}$ with the highest $T_C$ at ambient pressure. *npj Quant. Mater.* **2025**, *10*, 20.

(48) Uchida, S.; Ido, T.; Takagi, H.; Arima, T.; Tokura, Y.; Tajima, S. Optical spectra of $La_{2-x}Sr_xCuO_4$: Effect of carrier doping on the electronic structure of the $CuO_2$ plane. *Phys. Rev. B* **1991**, *43*, 7942–7954.

(49) Timusk, T. Infrared properties of exotic superconductors. *Physica C: Supercond.* **1999**, *317-318*, 18–29.

(50) Basov, D. N.; Timusk, T. Electrodynamics of high-$T_C$ superconductors. *Rev. Mod. Phys.* **2005**, *77*, 721–779.

(51) Padilla, W. J.; Lee, Y. S.; Dumm, M.; Blumberg, G.; Ono, S.; Segawa, K.; Komiya, S.; Ando, Y.; Basov, D. N. Constant effective mass across the phase diagram of high-$T_C$ cuprates. *Phys. Rev. B* **2005**, *72*, 060511(R).

(52) Zaanen, J.; Gunnarsson, O. Charged magnetic domain lines and the magnetism of high-$T_c$ oxides. *Phys. Rev. B* **1989**, *40*, 7391(R)–7394(R). (53)

(53) Machida, K. Magnetism in $La_2CuO_4$ based compounds. *Physica C: Superconductivit*y **1989**, *158*, 192–196.

(54) Kato, M.; Machida, K.; Nakanishi, H.; Fujita, M. Soliton Lattice Modulation of Incommensurate Spin Density Wave in Two Dimensional Hubbard Model - A Mean Field Study. *Journal of the Physical Society of Japan* **1990**, *59*, 1047–1058.

(55) Tranquada, J. M.; Axe, J. D.; Ichikawa, N.; Nakamura, Y.; Uchida, S.; Nachumi, B. Neutron-scattering study of stripe-phase order of holes and spins in $La_{1.48}Nd_{0.4}Sr_{0.12}CuO_4$. *Phys. Rev.*

*B* **1996**, *54*, 7489–7499.

(56) Bianconi, A.; Saini, N. L.; Lanzara, A.; Missori, M.; Rossetti, T.; Oyanagi, H.; Yamaguchi, H.; Oka, K.; Ito, T. Determination of the Local Lattice Distortions in the $CuO_2$ Plane of $La_{1.85}Sr_{0.15}CuO_4$. *Phys. Rev. Lett.* **1996**, *76*, 3412–3415.

(57) Emery, V. J.; Kivelson, S. A.; Zachar, O. Spin-gap proximity effect mechanism of high-temperature superconductivity. *Phys. Rev. B* **1997**, *56*, 6120–6147.

(58) Haase, J.; Slichter, C. P.; Stern, R.; Milling, C. T.; Hinks, D. G. NMR Evidence for Spatial Modulations in the Cuprates. *J. Supercond.* **2000**, *13*, 723–726.

(59) Tranquada, J. M.; Woo, H.; Perring, T. G.; Goka, H.; Gu, G. D.; Xu, G.; Fujita, M.; Yamada, K. Quantum magnetic excitations from stripes in copper oxide superconductors. *Nature* **2004**, *429*, 534–538.

(60) Lee, H.; Kuo, C.-T.; Fujita, M.; Kao, C.-C.; Lee, J.-S. Superconductivity Reinforces Charge Density-Wave Phase Coherence across Cuprates. *Phys. Rev. Lett.* **2026**, *136*, 186502.

(61) Wolf, E. L.; Arnold, G. B.; Gurvitch, M. A.; Zasadzinski, J. F. *Josephson Junctions:History, Devices, and Applications*; CRC Press, 2017.

(62) Photopoulos, R.; Frésard, R. A 3D Tight-Binding Model for La-Based Cuprate Superconductors. *Ann. Phys.* **2019**, *531*, 1900177.

(63) Sebastian, S. E.; Harrison, N.; Goddard, P. A.; Altarawneh, M. M.; Mielke, C. H.; Liang, R.; Bonn, D. A.; Hardy, W. N.; Andersen, O. K.; Lonzarich, G. G. Compensated electron and hole pockets in an underdoped high-$T_c$ superconductor. *Phys. Rev. B* **2010**, *81*, 214524.

(64) Norman, M. R. Fermi-surface reconstruction and the origin of high-temperature superconductivity. *Physics* **2010**, *3*, 86.

(65) Xiang, H.; Lee, C.; Koo, H.-J.; Gong, X.; Whangbo, M.-H. Magnetic properties and energy-

mapping analysis. *Dalton Trans.* **2013**, *42*, 823–853.

(66) Whangbo, M.-H.; Torardi, C. C. Hole Density Dependence of the Critical temperature and Coupling Constant in the Cuprate Superconductors. *Science* **1990**, *249*, 1143–1146.

(67) Whangbo, M.-H.; Torardi, C. C. Structure-Property Correlations in Cuprate Superconductors. *Acc. Chem. Res.* **1991**, *24*, 127–133.

(68) Albright, T. A.; Burdett, J. K.; Whangbo, M.-H. *Orbital Interactions in Chemistry* 2nd ed.; John Wiley & Sons: New York, 2013.

(69) Vinograd, I.; Zhou, R.; Hirata, M.; Wu, T.; Mayaffre, H.; Krämer, S.; Liang, R.; Hardy, W. N.; Bonn, D. A.; Julien, M.-H. Locally commensurate charge-density wave with three-unit-cell periodicity in $YBa_2Cu_3O_y$. *Nature Commun.* **2021**, *12*, 3274.

(70) Miao, H.; Fumagalli, R.; Rossi, M.; Lorenzana, J.; Seibold, G.; Yakhou-Harris, F.; Kummer, K.; Brookes, N. B.; Gu, G. D.; Braicovich, L.; Ghiringhelli, G.; Dean, M. P. M. Formation of Incommensurate Charge Density Waves in Cuprates. *Phys. Rev. X* **2019**, *9*, 031042.

(71) Tranquada, J. M. Cuprate superconductors as viewed through a striped lens. *Adv. Phys.* **2020**, *69*, 437–509.

(72) Hayden, S. M.; Tranquada, J. M. Charge Correlations in Cuprate Superconductors. *Ann. Rev. Condens. Matter Phys.* **2024**, *15*, 215–235.

(73) Fujita, M.; Goka, H.; Yamada, K.; Tranquada, J. M.; Regnault, L. P. Stripe order, depinning, and fluctuations in $La_{1.875}Ba_{0.125}CuO_4$ and $La_{1.875}Ba_{0.075}Sr_{0.050}CuO_4$. *Phys. Rev. B* **2004**, *70*, 104517.

(74) Hücker, M.; v. Zimmermann, M.; Gu, G. D.; Xu, Z. J.; Wen, J. S.; Xu, G. t.; Kang, H. J.; Zheludev, A.; Tranquada, J. M. Stripe order in superconducting $La_{2-x}Ba_xCuO_4$ ($0.095 \leq x \leq 0.155$). *Phys. Rev. B* **2011**, *83*, 104506.

(75) Ma, Q.; Rule, K. C.; Cronkwright, Z. W.; Dragomir, M.; Mitchell, G.; Smith, E. M.; Chi, S.; Kolesnikov, A. I.; Stone, M. B. et al. Parallel spin stripes and their coexistence with superconducting ground states at optimal and high doping in $La_{1.6-x}Nd_{0.4}Sr_xCuO_4$. *Phys. Rev. Res.* **2021**, *3*, 023151.

(76) Gupta, N. K.; McMahon, C.; Sutarto, R.; Shi, T.; Gong, R.; Wei, H.  I.; Shen, K. M.; He, F. p.; Ma, Q.; Dragomir, M.; Gaulin, B. D.; Hawthorn, D. G. Vanishing nematic order beyond the pseudogap phase in overdoped cuprate superconductors. *Proceedings of the National Academy of Sciences* **2021**, *118*, e2106881118.

(77) Lee, S.; Huang, E. W.; Johnson, T. A.; Guo, X.; Husain, A. A.; Mitrano, M.; Lu, K.; Zakrzewski, O.; de la Cruz, C.; MacDougall, G. J.; Chiang, T. C.; Fradkin, E.; Abbamonte, P. Generic character of charge and spin density waves in superconducting cuprates. *Proc. Nat. Acad. Sci.* **2022**, *119*, e2119429119.

(78) Wan, X.; Maier, T. A.; Savrasov, S. Y. Calculated magnetic exchange interactions in high-temperature superconductors. *Phys. Rev. B* **2009**, *79*, 155114.

(79) Ofer, R.; Bazalitsky, G.; Kanigel, A.; Keren, A.; Auerbach, A.; Lord, J. S.; Amato, A. Magnetic analog of the isotope effect in cuprates. *Phys. Rev. B* **2006**, *74*, 220508(R).

(80) Coldea, R.; Hayden, S.  M.; Aeppli, G.; Perring, T. G.; Frost, C. D.; Mason, T. E.; Cheong, S.-W.; Fisk, Z. Spin Waves and Electronic Interactions in $La_2CuO_4$. *Phys. Rev. Lett.* **2001**, *86*, 5377–5380.

(81) Reznik, D.; Bourges, P.; Pintschovius, L.; Endoh, Y.; Sidis, Y.; Masui, T.; Tajima, S. Dispersion of Magnetic Excitations in Optimally Doped Superconducting $YBa_2Cu_3O_{6.95}$. *Phys. Rev. Lett.* **2004**, *93*, 207003.

(82) Hashimoto, M.; Vishik, I. M.; He, R.-H.; Devereaux, T. P.; Shen, Z.-X. Energy gaps in high-

transition-temperature cuprate superconductors. *Nature Phys.* **2014**, *10*, 483–495.

(83) Miyake, T.; Imaizumi, T.; Iguchi, I. *d*-wave anisotropy and coexistence of the superconducting gap and quasiparticle excitations in $YBa_2Cu_3O_{7-\delta}$ Josephson junctions. *Phys. Rev. B* **2003**, *68*, 214520. (84)

(84) Tallon, J. L.; Storey, J. G. Thermodynamics of the pseudogap in cuprates. *Front. Phys.* **2022**, *10*, 1030616.

(85) Niu, J.; Larrazabal, M. O.; Gozlinski, T.; Sato, Y.; Bastiaans, K. M.; Ben-schop, T.; Ge, J.-F.; Blanter, Y. M.; Gu, G.; Swart, I.; Allan, M. P. Equivalence of pseudogap and pairing energy in a cuprate high-temperature superconductor. 2024; https://arxiv.org/abs/2409.15928.

(86) Chen, Q.; Moskal, A.; Wang, Y.; McNiven, B. D. E.; Aczel, A. A.; Tian, W.; Gaulin, B. D. Competing pair density wave and uniform *d*-wave superconductivity in phase-separated 214 cuprates at the 1/8 anomaly. *Phys. Rev. B* **2025**, *112*, 174506.

(87) Tranquada, J. M.; Dean, M. P. M.; Li, Q. Superconductivity from Charge Order in Cuprates. *J. Phys. Soc. Jpn* **2021**, *90*, 111002.

(88) Choubey, P.; Joo, S. H.; Fujita, K.; Du, Z.; Edkins, S. D.; Hamidian, M. H.; Eisaki, H.; Uchida, S.-i. A.; Mackenzie, A. P.; Lee, J.; Davis, J. C. S.; Hirschfeld, P. J. Atomic-scale electronic structure of the cuprate pair density wave state coexisting with superconductivity. *Proc. Nat. Acad. Sci.* **2020**, *117*, 14805–14811.

(89) Patel, A. A.; Lunts, P.; Albergo, M. S. Strange Metals and Planckian Transport in a Gapless Phase from Spatially Random Interactions. *Phys. Rev. X* **2025**, *15*, 031064.

TOC figure

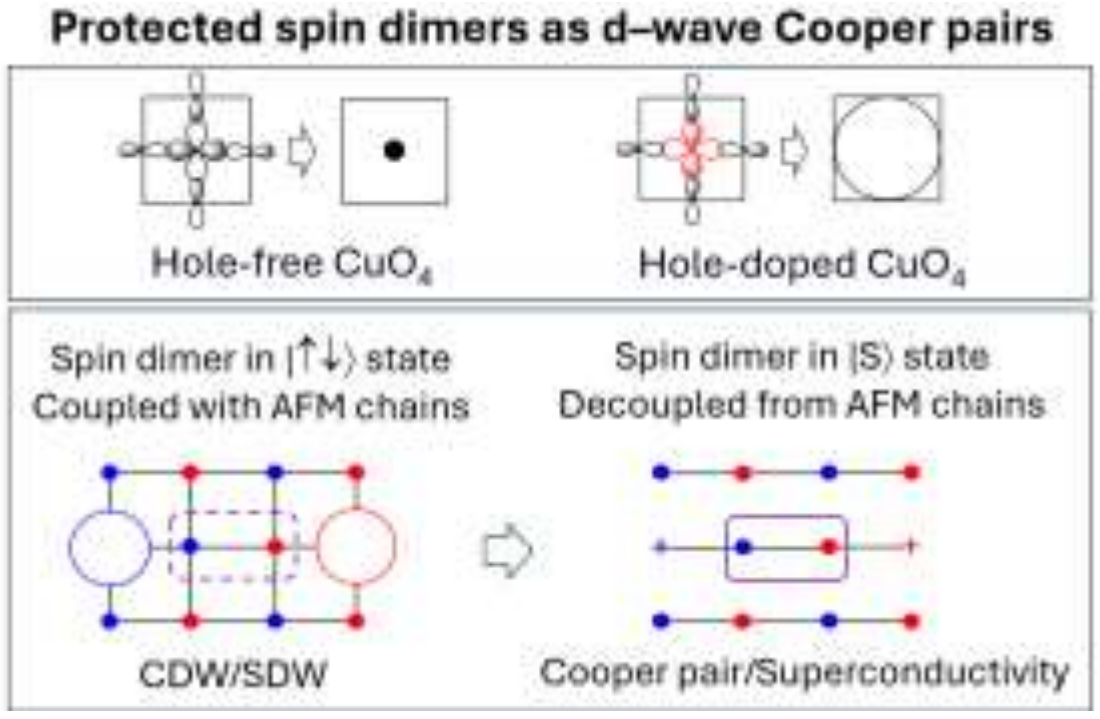

Protected spin dimers as d–wave Cooper pairs
Hole-free CuO4
Hole-doped CuO4
Spin dimer in |↑↓⟩ state
Coupled with AFM chains
Spin dimer in |S⟩ state
Decoupled from AFM chains
CDW/SDW
Cooper pair/Superconductivity